\documentclass[twocolumn,10pt]{IEEEtran}

\usepackage[acronym]{glossaries}
\newacronym{iid}{i.i.d.}{independent and identically distributed}
\newacronym{mimo}{MIMO}{multiple-input multiple-output}
\newacronym{miso}{MISO}{multiple-input single-output}
\newacronym{adc}{ADC}{analog-to-digital converter}
\newacronym{dac}{DAC}{digital-to-analog converter}
\newacronym{rf}{RF}{radio-frequency}
\newacronym{mmwave}{mmWave}{millimeter wave}
\newacronym{dft}{DFT}{discrete Fourier transform}
\newacronym{doa}{DoA}{direction-of-arrival}
\newacronym{awgn}{AWGN}{additive white Gaussian noise}
\newacronym{snr}{SNR}{signal-to-noise ratio}
\newacronym{sqnr}{SQNR}{signal-to-quantization-noise ratio}
\newacronym{qpsk}{QPSK}{quadrature phase shift keying}
\newacronym{ber}{BER}{bit error rate}
\newacronym{csi}{CSI}{channel state information}
\newacronym{em}{EM}{electromagnetic}

\newacronym{espar}{ESPAR}{electronically steerable parasitic array radiator}
\newacronym{lma}{LMA}{load modulated array}

\newacronym{mse}{MSE}{mean square error}
\newacronym{mmse}{MMSE}{minimum mean square error}
\newacronym{lmmse}{LMMSE}{linear minimum mean square error}
\newacronym{gls}{GLS}{generalized least squares}
\newacronym{gmf}{GMF}{generalized matched filtering}
\newacronym{rls}{RLS}{regularized least squares}
\newacronym{ols}{OLS}{ordinary least squares}
\newacronym{omf}{OMF}{ordinary matched filtering}

\newacronym{zf}{ZF}{zero-forcing}
\newacronym{ls}{LS}{least squares}
\newacronym{mf}{MF}{matched filtering}
\newacronym{r-zfbf}{R-ZFBF}{regularized zero-forcing beamforming}
\newacronym{zfbf}{ZFBF}{zero-forcing beamforming}
\newacronym{mbf}{MBF}{matched beamforming}

\newacronym{ris}{RIS}{reconfigurable intelligent surface}
\newacronym{bd-ris}{BD-RIS}{beyond-diagonal RIS}
\newacronym{sim}{SIM}{stacked intelligent metasurface}
\newacronym{dsp}{DSP}{digital signal processor}
\newacronym{asp}{ASP}{analog signal processing}

\newacronym{milac}{MiLAC}{microwave linear analog computer}

\newacronym{isa}{ISA}{instruction set architecture}

\newacronym{nn}{NN}{neural network}

\newacronym{cpu}{CPU}{central processing unit}

\newacronym{dma}{DMA}{dynamic metasurface antenna}
\newacronym{pwm}{PWM}{pulse-width modulation}

\usepackage{amsmath,amssymb,amsthm}
\usepackage{graphicx}
\usepackage{comment}
\usepackage{xcolor}
\usepackage{subfigure}
\usepackage{cases}
\usepackage{booktabs}
\usepackage{algorithm}
\usepackage{algorithmic}
\usepackage{multirow}

\newtheorem{remark}{Remark}

\begin{document}
\bstctlcite{BSTcontrol}

\title{Analog Computing for Signal Processing\\and Communications -- Part~II:\\Toward Gigantic MIMO Beamforming}

\author{Matteo~Nerini,~\IEEEmembership{Member,~IEEE},
        Bruno~Clerckx,~\IEEEmembership{Fellow,~IEEE}

\thanks{Part of this work has been accepted by the 2025 IEEE Global Communications Conference (GLOBECOM) \cite{ner25-2}. (Corresponding author: Bruno Clerckx.)}
\thanks{This work has been partially supported by UKRI grant EP/Y004086/1, EP/X040569/1, EP/Y037197/1, EP/X04047X/1, and EP/Y037243/1.}
\thanks{Matteo Nerini and Bruno Clerckx are with the Department of Electrical and Electronic Engineering, Imperial College London, London SW7 2AZ, U.K. (e-mail: \{m.nerini20, b.clerckx\}@imperial.ac.uk).}
\thanks{Bruno Clerckx is also with Kyung Hee University, Seoul, Korea.}}

\maketitle

\begin{abstract}
Analog-domain operations offer a promising solution to accelerating signal processing and enabling future \gls{mimo} communications with thousands of antennas.
In Part~I of this paper, we have introduced a \gls{milac} as an analog computer that processes microwave signals linearly, demonstrating its potential to reduce the computational complexity of specific signal processing tasks.
In Part~II of this paper, we extend these benefits to wireless communications, showcasing how \gls{milac} enables gigantic \gls{mimo} beamforming entirely in the analog domain.
\Gls{milac}-aided beamforming enables the maximum flexibility and performance of digital beamforming, while significantly reducing hardware costs by minimizing the number of \gls{rf} chains and only relying on low-resolution \glspl{adc} and \glspl{dac}.
In addition, it eliminates per-symbol operations by completely avoiding digital-domain processing and remarkably reduces the computational complexity of \gls{zf}, which scales quadratically with the number of antennas instead of cubically.
It also processes signals with fixed matrices, e.g., the \gls{dft}, directly in the analog domain.
Numerical results show that it can perform \gls{zf} and \gls{dft} with a computational complexity reduction of up to $1.5\times10^4$ and $4.0\times10^7$ times, respectively, compared to digital beamforming.
\end{abstract}

\glsresetall


\section{Introduction}

In Part~I of this paper \cite{ner25-3}, we have reviewed the growing interest in analog computing, driven by advancements in several technologies such as optical systems, metamaterials, and resistive memory arrays.
Digital computing, despite its dominance, faces fundamental limitations in speed and power consumption, mainly due to the reliance on \glspl{adc} and \glspl{dac}.
Analog computing addresses these challenges by enabling energy-efficient and highly parallelized computations at the speed of light.
While Part I has focused on how analog computing can be leveraged for signal processing operations, Part II explores how these principles can be applied to achieve efficient and flexible beamforming in the analog domain, paving the way for future wireless communications.

As we are witnessing a revival of interest in analog operations within the field of signal processing, a similar shift from digital to analog has emerged in wireless communications.
This shift is driven by the growing number of antennas needed to support higher data rates and simultaneously serve a larger number of users.
While massive \gls{mimo} systems typically deploy 64 or more antennas \cite{lar14,bjo16}, this number could reach several thousands in ultra-massive or gigantic \gls{mimo} systems, which have been recently proposed to meet the stringent requirements of 6G networks \cite{bjo24}.
With such a large number of antennas, digital beamforming faces two critical challenges in terms of hardware and computational complexity.
First, digital beamforming typically requires a dedicated \gls{rf} chain per antenna element, each including \glspl{adc}/\glspl{dac} and mixers, which are costly and power-hungry components.
Second, as the number of antennas increases, the data volumes to be processed in real-time grow accordingly.
For example, to precode the transmitted symbol vector at the transmitter (or combine the received symbol vector at the receiver) a matrix-vector product is required on a per-symbol basis.
These two challenges result in high hardware costs and computational complexity when digital beamforming is applied in massive or gigantic \gls{mimo} systems.
For these reasons, several alternatives to digital beamforming have been explored to enable beamforming operations fully or partially in the analog domain.

To address the high hardware cost associated with digital beamforming, alternative strategies such as analog beamforming and hybrid analog-digital beamforming have been proposed.
Analog beamforming steers the beam pattern by solely requiring a single \gls{rf} chain \cite{sun14}.
This is achieved by connecting the \gls{rf} chain to the antennas via phase shifters or time delay elements, which impose a constant modulus constraint on the beamforming weights.
Thus, while analog beamforming offers low hardware complexity, it is characterized by limited flexibility.
To balance hardware cost and flexibility by bridging analog and digital beamforming, hybrid analog-digital beamforming techniques have been widely investigated \cite{sun14,aya14,soh16,mol17,ahm18,gon20}.
In hybrid beamforming, the beamforming process is divided into two stages.
First, the transmitted symbols are digitally processed in the baseband domain with a low-dimensional digital beamforming matrix, and fed to a reduced number of \gls{rf} chains.
Second, the \gls{rf} chains are connected to the transmitting antennas through a network of phase shifters, operating analog beamforming in the \gls{rf} domain.
The reversed process is applied at the receiver.

Another method to efficiently perform beamforming in the analog domain is given by \gls{ris} \cite{dir20,wu21-2}.
This technology has recently emerged enabling the control of the wireless channel through surfaces made of multiple passive elements with reconfigurable scattering properties.
Beamforming in the analog domain can be achieved through \gls{ris} by deploying a \gls{ris} in close proximity to an active transceiving device equipped with a reduced number of antennas.
In this way, the effective radiation pattern of the active device can be optimized by reconfiguring the \gls{ris} elements \cite{jam21,you22,hua23}.
Two extensions of this technology have been proposed to achieve additional flexibility over conventional \gls{ris}.
First, \gls{bd-ris} include more general \gls{ris} architectures, characterized by a scattering matrix not restricted to be diagonal \cite{she20,li22-1,ner23-1}.
By deploying a \gls{bd-ris} close to an active device, its additional flexibility over conventional \gls{ris} in manipulating the radiation pattern can enable even higher performance \cite{mis24}.
Second, \gls{sim} technology considers multiple stacked transmissive \glspl{ris} deployed close to the active device \cite{an23,an24a,ner24}.
The additional flexibility provided by the multiple layers has shown increased performance over just deploying a single conventional \gls{ris}.
Reconfigurable metasurfaces similar to those in \gls{ris} and \gls{sim} have been proposed to enable \glspl{dma}, and perform beamforming and combining in the analog domain \cite{shl19,wil22,pro25}.

In the reviewed beamforming strategies operating in the analog domain, the number of \gls{rf} chains is commonly lower bounded by the number of symbols, or streams, transmitted in parallel.
Nevertheless, other beamforming strategies have been developed to also perform symbol modulation in the analog domain, allowing the transmission of multiple symbols with a single \gls{rf} chain.
First, in \glspl{espar}, the symbols are modulated and the radiation pattern reconfigured by adjusting the mutual coupling between one active antenna element and multiple ``parasitic'' elements having tunable loads \cite{kaw05,han13,zha23}.
Second, in \glspl{lma}, a single \gls{rf} source is connected to multiple antennas through tunable impedance components, which are reconfigured to achieve symbol modulation \cite{mul14,sed16}.
Third, symbol modulation can be performed in the analog domain by employing a \gls{ris} deployed close to an active antenna transmitting a carrier signal \cite{tan20a,tan20b}.
These three strategies further highlight the potential of leveraging analog-domain beamforming to reduce the hardware cost of transceiver devices.

In addition to alleviating the hardware cost, analog-domain strategies have been also explored to reduce the computational complexity of future wireless networks by performing specific mathematical operations in the analog domain.
To this end, in-memory computing has been proposed to accelerate ridge regression with applications to massive \gls{mimo} \cite{man22,man23c,wan23}.
While in-memory computing approaches can efficiently process baseband signals, other strategies have emerged to directly perform analog processing at \gls{rf}.
\gls{ris} and \gls{sim} have been used to compute the \gls{dft} of the incident signal in the analog domain, which is particularly beneficial for \gls{doa} estimation \cite{oma25b,joy25,an24b}.
Moreover, over-the-air computing emerged as a form of analog computing that leverages the superposition principle characterizing the \gls{em} propagation over the wireless channels.
Over-the-air computing allows efficient computation of the so-called nomographic functions \cite{gol13,zhu19,zhu21}, with impactful applications on federated learning \cite{yan20}.

Numerous technologies have been proposed to reduce the hardware cost of future wireless transceivers through analog beamforming \cite{sun14}-\cite{tan20b} and analog computing \cite{man22}-\cite{yan20}.
However, operating in the analog domain typically decreases the flexibility compared to digital beamforming, with a consequent reduction in performance.
Thus, whether it is possible to maintain maximum performance while simultaneously minimizing hardware cost and computational complexity through analog computing remains an open question.
In Part~II of this paper, we address this gap by introducing a novel beamforming strategy enabled by \glspl{milac}, denoted as \gls{milac}-aided beamforming.
This beamforming strategy maintains the same flexibility as digital beamforming while operating entirely in the analog domain.
Because of its fully analog nature, it minimizes the number of \gls{rf} chains and relies only on low-resolution \glspl{adc}/\glspl{dac}, reducing the hardware cost.
By leveraging the computational capabilities of \glspl{milac}, it also minimizes the computational complexity required for beamforming.
More in detail, the contributions of Part~II of this paper are as follows.

\textit{First}, we demonstrate that a \gls{milac} can efficiently process microwave signals to compute five mathematical operations of practical interest in \gls{mimo} communications, such as \gls{ls} and \gls{mf}.
To this end, we first show how these five operations can be seen as special cases of the \gls{lmmse} estimator for linear observation processes, and then discuss how the microwave network of the \gls{milac} needs to be reconfigured to compute them.
These operations can be computed with a \gls{milac} with significantly reduced computational complexity compared to conventional digital computing.

\textit{Second}, we propose \gls{milac}-aided beamforming as a new beamforming strategy enabled by \gls{milac} scalable to gigantic \gls{mimo} systems.
\Gls{milac}-aided beamforming can perform precoding at the transmitter and combining at the receiver, directly operating with microwave signals, and maintaining the same flexibility as digital beamforming.
While having the maximum flexibility, \gls{milac}-aided beamforming allows us to perform specific operations in the analog domain, such as \gls{zf} precoding at the transmitter and \gls{zf} combining at the receiver, with important gains in terms of computational complexity.
In addition, it allows to precode or combine a signal with a fixed matrix, such as the \gls{dft} matrix, entirely in the analog domain, requiring no digital operations.

\textit{Third}, we compare \gls{milac}-aided beamforming with existing beamforming strategies, namely digital, analog, hybrid, \gls{ris}-aided, and \gls{sim}-aided beamforming, in terms of flexibility of the beamforming matrix, hardware complexity, and computational complexity.
As a result of this comparison, we discuss five benefits of \gls{milac}-aided beamforming:
\textit{1)} it enables the maximum flexibility,
\textit{2)} it requires the minimum number of \gls{rf} chains,
\textit{3)} it only needs low-resolution \glspl{adc}/\glspl{dac},
\textit{4)} it does not require any computation on a per-symbol basis, and
\textit{5)} it requires the minimum number of computations per coherence block for \gls{zf}.

\textit{Fourth}, we provide numerical results to assess the performance of \gls{milac}-aided beamforming in terms of sum rate and \gls{ber}, confirming that it can achieve the same performance as digital beamforming.
We also assess the benefits of \gls{milac}-aided beamforming over digital beamforming in terms of computational complexity, showing that it can perform \gls{zf}, \gls{mf}, and the \gls{dft} with a gain in computational complexity of $1.5\times10^4$, $2.0\times10^2$, and $4.0\times10^7$ times, respectively.
Thus, in a gigantic \gls{mimo} system with thousands of antennas, \gls{milac}-aided beamforming achieves the same performance as digital \gls{zf} beamforming at just $1/15000$th of the computational cost.
From a different perspective, a 4096-antenna \gls{milac} serving 4096 users via \gls{zf} beamforming requires the same computational complexity as a 256-antenna digital transmitter serving 256 users.
This makes \gls{milac}-aided beamforming a game changer for gigantic \gls{mimo}, opening the door to radically new transceiver architectures that can massively scale the number of antennas and users.

\textit{Organization}:
In Section~\ref{sec:milac}, we briefly review the concept of \gls{milac}, as introduced in Part~I of this paper.
In Section~\ref{sec:special-cases}, we derive five special cases of the \gls{lmmse} estimator and show how they can be efficiently computed by a \gls{milac}.
In Section~\ref{sec:gen-analog-bf}, we propose \gls{milac}-aided beamforming as a novel beamforming strategy enabled by a \gls{milac}.
In Section~\ref{sec:comparison}, we compare \gls{milac}-aided beamforming with existing beamforming strategies.
In Section~\ref{sec:results}, we provide numerical results to evaluate the performance and computational complexity of \gls{milac}-aided beamforming.
Finally, Section~\ref{sec:conclusion} concludes this paper.

\textit{Notation}:
Please, refer to the notation introduced in Section~I of Part~I of this paper \cite{ner25-3}.

\begin{figure}[t]
\centering
\includegraphics[width=0.42\textwidth]{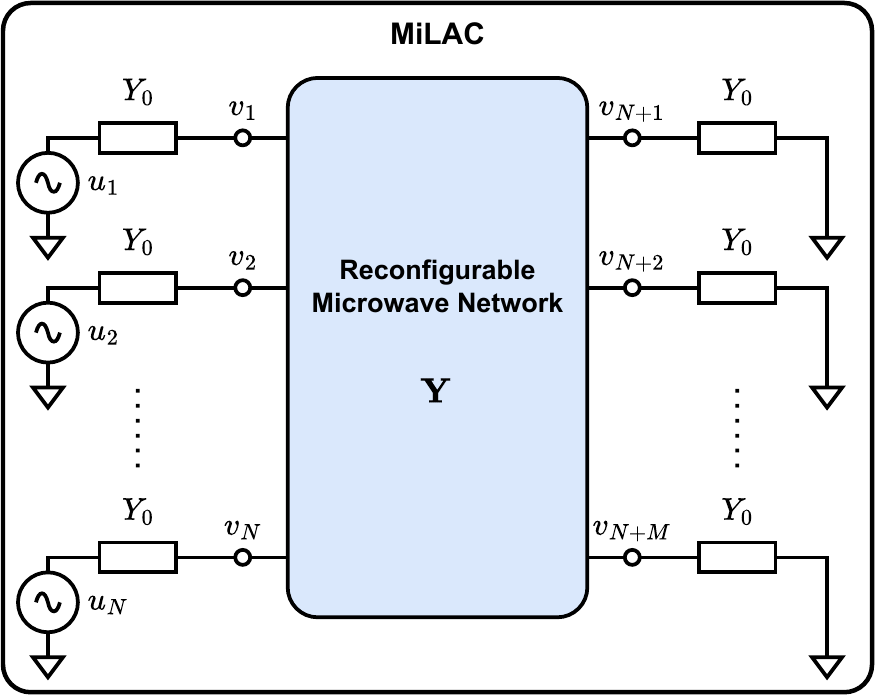}
\caption{Representation of a $P$-port MiLAC with input on $N$ ports, where $P=N+M$.}
\label{fig:ac}
\end{figure}

\section{Microwave Linear Analog Computer}
\label{sec:milac}

In this section, we briefly review the concept of \gls{milac}, defined as an analog computer that linearly processes microwave signals, as introduced in Part~I of this paper \cite{ner25-3}.
A \gls{milac} can be seen as a $P$-port reconfigurable microwave network, which can be characterized through its impedance, admittance, or scattering matrix according to multiport network theory.
Since these three representations are equivalent, we chose the admittance matrix $\mathbf{Y}\in\mathbb{C}^{P\times P}$ since its entries are closely related to the tunable components in the MiLAC, as seen in Part~I.
We assume this microwave network to be made of $P^2$ tunable admittance components interconnecting each port to ground and to all the other ports.
The admittance component interconnecting port $k$ to ground is denoted as $Y_{k,k}\in\mathbb{C}$, for $k=1,\ldots,P$, while the admittance component interconnecting port $k$ to port $i$ is denoted as $Y_{i,k}\in\mathbb{C}$, for $\forall i\neq k$.
Depending on the admittance values $\{Y_{i,k}\}_{i,k=1}^{P}$, the entries of the admittance matrix $\mathbf{Y}$ are give by
\begin{equation}
\left[\mathbf{Y}\right]_{i,k}=
\begin{cases}
-Y_{i,k} & i\neq k\\
\sum_{p=1}^PY_{p,k} & i=k
\end{cases},\label{eq:Yik-entry}
\end{equation}
for $i,k=1,\ldots,P$, as derived in Part~I.
The relationship in \eqref{eq:Yik-entry} shows that if the $P^2$ tunable admittance components are arbitrarily reconfigurable, then the admittance matrix $\mathbf{Y}$ can also be arbitrarily reconfigured.

The input signal $\mathbf{u}=[u_1,\ldots,u_N]^T\in\mathbb{C}^{N\times 1}$ is applied by $N$ voltage generators with series admittance $Y_0$ to the first $N$ ports of the \gls{milac}, with $N\leq P$, as represented in Fig.~\ref{fig:ac}.
The output signal is read on all $P$ ports, where $P=N+M$ with $M$ denoting the number of ports with no input, and is denoted by $\mathbf{v}=[v_1,\ldots,v_P]^T\in\mathbb{C}^{P\times 1}$, as shown in Fig.~\ref{fig:ac}.
For convenience of representation, we introduce the output vectors $\mathbf{v}_1$ and $\mathbf{v}_2$ as $\mathbf{v}_1=[v_1,\ldots,v_N]^T\in\mathbb{C}^{N\times 1}$ and $\mathbf{v}_2=[v_{N+1},\ldots,v_p]^T\in\mathbb{C}^{M\times 1}$, such that $\mathbf{v}=[\mathbf{v}_1^T,\mathbf{v}_2^T]^T$.
We consider a \gls{milac} with $M>0$ ports with no input in the following.

It has been shown in Part~I that the output vectors $\mathbf{v}_1$ and $\mathbf{v}_2$ can be expressed as a function of the admittance matrix $\mathbf{Y}$ and the input $\mathbf{u}$ by introducing a matrix $\mathbf{P}\in\mathbb{C}^{P\times P}$ as
\begin{equation}
\mathbf{P}=\frac{\mathbf{Y}}{Y_0}+\mathbf{I}_{P},\label{eq:P}
\end{equation}
partitioned as
\begin{equation}
\mathbf{P}=
\begin{bmatrix}
\mathbf{P}_{11} & \mathbf{P}_{12}\\
\mathbf{P}_{21} & \mathbf{P}_{22}
\end{bmatrix},
\end{equation}
where $\mathbf{P}_{11}\in\mathbb{C}^{N\times N}$, $\mathbf{P}_{12}\in\mathbb{C}^{N\times M}$, $\mathbf{P}_{21}\in\mathbb{C}^{M\times N}$, and $\mathbf{P}_{22}\in\mathbb{C}^{M\times M}$.
Assuming $\mathbf{P}$ to be invertible, with $\mathbf{P}_{11}$ or $\mathbf{P}_{22}$ invertible, $\mathbf{v}_1$ and $\mathbf{v}_2$ are expressed depending on the invertibility of $\mathbf{P}_{11}$ and $\mathbf{P}_{22}$ as follows:
\begin{enumerate}
\item If $\mathbf{P}_{11}$ is invertible, we have
\begin{multline}
\mathbf{v}_1=\Big(\mathbf{P}_{11}^{-1}-\mathbf{P}_{11}^{-1}\mathbf{P}_{12}\Big.\\
\left.\times\left(\mathbf{P}_{21}\mathbf{P}_{11}^{-1}\mathbf{P}_{12}-\mathbf{P}_{22}\right)^{-1}\mathbf{P}_{21}\mathbf{P}_{11}^{-1}\right)\mathbf{u},\label{eq:v1-1}
\end{multline}
\begin{equation}
\mathbf{ v}_2=\left(\mathbf{P}_{21}\mathbf{P}_{11}^{-1}\mathbf{P}_{12}-\mathbf{P}_{22}\right)^{-1}\mathbf{P}_{21}\mathbf{P}_{11}^{-1}\mathbf{u}.\label{eq:v2-1}
\end{equation}
\item If $\mathbf{P}_{22}$ is invertible, we have
\begin{equation}
\mathbf{v}_1=-\left(\mathbf{P}_{12}\mathbf{P}_{22}^{-1}\mathbf{P}_{21}-\mathbf{P}_{11}\right)^{-1}\mathbf{u},\label{eq:v1-2}
\end{equation}
\begin{equation}
\mathbf{v}_2=\mathbf{P}_{22}^{-1}\mathbf{P}_{21}\left(\mathbf{P}_{12}\mathbf{P}_{22}^{-1}\mathbf{P}_{21}-\mathbf{P}_{11}\right)^{-1}\mathbf{u}.\label{eq:v2-2}
\end{equation}
\item If $\mathbf{P}_{11}$ and $\mathbf{P}_{22}$ are both invertible, \eqref{eq:v1-1}-\eqref{eq:v2-1} and \eqref{eq:v1-2}-\eqref{eq:v2-2} are equivalent.
\end{enumerate}
Note that \eqref{eq:v1-1}-\eqref{eq:v2-1} and \eqref{eq:v1-2}-\eqref{eq:v2-2} express the output vectors $\mathbf{v}_1$ and $\mathbf{v}_2$ as a function of the input $\mathbf{u}$ and the blocks of $\mathbf{P}$, which are directly related to $\mathbf{Y}$ by \eqref{eq:P}.
The expressions \eqref{eq:v1-1}-\eqref{eq:v2-1} and \eqref{eq:v1-2}-\eqref{eq:v2-2} can be computed by a \gls{milac} given any blocks $\mathbf{P}_{11}$, $\mathbf{P}_{12}$, $\mathbf{P}_{21}$, and $\mathbf{P}_{22}$ by setting the tunable admittance components $\{Y_{i,k}\}_{i,k=1}^{P}$ as a function of any arbitrary $\mathbf{P}$ as
\begin{equation}
Y_{i,k}=
\begin{cases}
-Y_0\left[\mathbf{P}\right]_{i,k} & i\neq k\\
Y_0\sum_{p=1}^P\left[\mathbf{P}\right]_{p,k}-Y_0 & i=k
\end{cases},\label{eq:Yik-component-P}
\end{equation}
for $i,k=1,\ldots,P$, as discussed in Part~I.
Remarkably, a \gls{milac} can efficiently compute \eqref{eq:v1-1}-\eqref{eq:v2-1} and \eqref{eq:v1-2}-\eqref{eq:v2-2} in the analog domain once the input signals are applied at the input ports, while they would require computationally expensive matrix-matrix products and matrix inversion operations if digitally computed.

\section{Analog Computing for the Special Cases of the LMMSE Estimator}
\label{sec:special-cases}

In Part~I of this paper, we have shown how a \gls{milac} can be used to compute the \gls{lmmse} estimator of a linear observation process with reduced computational complexity.
In this section, we derive and discuss five special cases of the \gls{lmmse} estimator for linear observation processes, of particular interest in wireless communications.
Then, we show how these five special cases can be efficiently computed by a \gls{milac}.

\subsection{LMMSE Estimator for Linear Observation Processes}
\label{sec:lmmse}

Consider a linear observation process $\mathbf{y}=\mathbf{H}\mathbf{x}+\mathbf{n}$, where $\mathbf{y}\in\mathbb{C}^{Y\times 1}$ is the observation vector, $\mathbf{H}\in\mathbb{C}^{Y\times X}$ is a known constant matrix, $\mathbf{x}\in\mathbb{C}^{X\times 1}$ is the unknown random vector with mean $\bar{\mathbf{x}}=\mathbf{0}_{X\times 1}$ and covariance matrix $\mathbf{C}_\mathbf{x}$, and $\mathbf{n}\in\mathbb{C}^{Y\times 1}$ is the random noise vector with mean $\bar{\mathbf{n}}=\mathbf{0}_{Y\times 1}$ and covariance matrix $\mathbf{C}_\mathbf{n}$.
We assume that $\mathbf{C}_{\mathbf{x}}$ and $\mathbf{C}_{\mathbf{n}}$ are known and invertible, and that the cross-covariance matrix of $\mathbf{x}$ and $\mathbf{n}$ is $\mathbf{C}_{\mathbf{x}\mathbf{n}}=\mathbf{0}_{X\times Y}$.
As discussed in part~I, the \gls{lmmse} estimator of $\mathbf{x}$ is equivalently given by
\begin{align}
\hat{\mathbf{x}}_{\textnormal{LMMSE},1}
&=\left(\mathbf{H}^H\mathbf{C}_{\mathbf{n}}^{-1}\mathbf{H}+\mathbf{C}_{\mathbf{x}}^{-1}\right)^{-1}\mathbf{H}^H\mathbf{C}_{\mathbf{n}}^{-1}\mathbf{y},\label{eq:LMMSE1}\\
\hat{\mathbf{x}}_{\textnormal{LMMSE},2}
&=\mathbf{C}_{\mathbf{x}}\mathbf{H}^H\left(\mathbf{H}\mathbf{C}_{\mathbf{x}}\mathbf{H}^H+\mathbf{C}_{\mathbf{n}}\right)^{-1}\mathbf{y}\label{eq:LMMSE2},
\end{align}
where we have $\hat{\mathbf{x}}_{\textnormal{LMMSE},1}=\hat{\mathbf{x}}_{\textnormal{LMMSE},2}$ when $\mathbf{H}^H\mathbf{C}_{\mathbf{n}}^{-1}\mathbf{H}+\mathbf{C}_{\mathbf{x}}^{-1}$ and $\mathbf{H}\mathbf{C}_{\mathbf{x}}\mathbf{H}^H+\mathbf{C}_{\mathbf{n}}$ are both invertible, which holds true in general.

\subsection{Special Cases of the LMMSE Estimator for Linear Observation Processes}

The \gls{lmmse} estimator in Section~\ref{sec:lmmse} has been obtained with no assumptions on the covariance matrices $\mathbf{C}_{\mathbf{x}}$ and $\mathbf{C}_\mathbf{n}$ beyond their invertibility.
With specific assumptions on $\mathbf{C}_{\mathbf{x}}$ and $\mathbf{C}_\mathbf{n}$, we can show that it includes five special cases of practical interest, as discussed in the following.

\subsubsection{Generalized Least Squares}

In \gls{gls}, also known as generalized linear regression, the variance on the a priori information on $\mathbf{x}$ is infinitely higher than on $\mathbf{n}$, i.e., $\Vert\mathbf{C}_{\mathbf{x}}\Vert_F\gg\Vert\mathbf{C}_{\mathbf{n}}\Vert_F$.
Thus, $\mathbf{C}_{\mathbf{x}}^{-1}$ is negligible in \eqref{eq:LMMSE1} and, equivalently, $\mathbf{C}_{\mathbf{n}}$ is negligible in \eqref{eq:LMMSE2}, yielding
\begin{align}
\hat{\mathbf{x}}_{\text{GLS},1}
&=\left(\mathbf{H}^H\mathbf{C}_{\mathbf{n}}^{-1}\mathbf{H}\right)^{-1}\mathbf{H}^H\mathbf{C}_{\mathbf{n}}^{-1}\mathbf{y},\label{eq:sc1}\\
\hat{\mathbf{x}}_{\text{GLS},2}
&=\mathbf{C}_{\mathbf{x}}\mathbf{H}^H\left(\mathbf{H}\mathbf{C}_{\mathbf{x}}\mathbf{H}^H\right)^{-1}\mathbf{y},
\end{align}
respectively.
Note that, depending on the dimensions of $\mathbf{H}$, only one matrix among $\mathbf{H}^H\mathbf{C}_{\mathbf{n}}^{-1}\mathbf{H}$ and $\mathbf{H}\mathbf{C}_{\mathbf{x}}\mathbf{H}^H$ may be invertible.
Thus, only one among the estimates $\hat{\mathbf{x}}_{\text{GLS},1}$ and $\hat{\mathbf{x}}_{\text{GLS},2}$ may exist.

\subsubsection{Generalized Matched Filtering}

Opposite to \gls{gls}, we can assume that the variance on the a priori information on $\mathbf{x}$ is infinitely smaller than on $\mathbf{n}$, i.e., $\Vert\mathbf{C}_{\mathbf{x}}\Vert_F\ll\Vert\mathbf{C}_{\mathbf{n}}\Vert_F$.
In this case, that we denote as \gls{gmf}, both \eqref{eq:LMMSE1} and \eqref{eq:LMMSE2} boil down to
\begin{equation}
\hat{\mathbf{x}}_{\text{GMF}}=\mathbf{C}_{\mathbf{x}}\mathbf{H}^H\mathbf{C}_{\mathbf{n}}^{-1}\mathbf{y}.
\end{equation}

\subsubsection{Regularized Least Squares}

In \gls{rls}, also known as ridge regression, the entries of $\mathbf{x}$ and $\mathbf{n}$ are uncorrelated and have equal variance, i.e., $\mathbf{C}_{\mathbf{x}}=\lambda_\mathbf{x}\mathbf{I}_X$ and $\mathbf{C}_{\mathbf{n}}=\lambda_\mathbf{n}\mathbf{I}_Y$, with $\lambda_\mathbf{x},\lambda_\mathbf{n}\in\mathbb{R}$ and $\lambda_\mathbf{x},\lambda_\mathbf{n}>0$.
Thus, by introducing $\lambda\in\mathbb{R}$ such that $\lambda=\lambda_\mathbf{n}/\lambda_\mathbf{x}$, \eqref{eq:LMMSE1} and \eqref{eq:LMMSE2} simplify as
\begin{align}
\hat{\mathbf{x}}_{\text{RLS},1}
&=\left(\mathbf{H}^H\mathbf{H}+\lambda\mathbf{I}_X\right)^{-1}\mathbf{H}^H\mathbf{y},\label{eq:RLS1}\\
\hat{\mathbf{x}}_{\text{RLS},2}
&=\mathbf{H}^H\left(\mathbf{H}\mathbf{H}^H+\lambda\mathbf{I}_Y\right)^{-1}\mathbf{y},\label{eq:RLS2}
\end{align}
respectively.
Interestingly, the expressions in \eqref{eq:RLS1} and \eqref{eq:RLS2} are widely used in \gls{mimo} wireless communications to realize the so-called \gls{r-zfbf} and the \gls{mmse} receiver, as effective linear transmission and reception techniques, respectively \cite[Chapters~7,~12]{cle13}.

\begin{table}[t]
\centering
\caption{Special cases of the LMMSE estimator and their assumptions.}
\begin{tabular}{@{}ll@{}}
\toprule
 & Assumptions on $\mathbf{C}_{\mathbf{x}}$ and $\mathbf{C}_{\mathbf{n}}$\\
\midrule
LMMSE & None\\
GLS & $\Vert\mathbf{C}_{\mathbf{x}}\Vert_F\gg\Vert\mathbf{C}_{\mathbf{n}}\Vert_F$\\
GMF & $\Vert\mathbf{C}_{\mathbf{x}}\Vert_F\ll\Vert\mathbf{C}_{\mathbf{n}}\Vert_F$\\
RLS & $\mathbf{C}_{\mathbf{x}}=\lambda_\mathbf{x}\mathbf{I}_X$, $\mathbf{C}_{\mathbf{n}}=\lambda_\mathbf{n}\mathbf{I}_Y$\\
OLS & $\mathbf{C}_{\mathbf{x}}=\lambda_\mathbf{x}\mathbf{I}_X$, $\mathbf{C}_{\mathbf{n}}=\lambda_\mathbf{n}\mathbf{I}_Y$, $\lambda_\mathbf{x}\gg\lambda_\mathbf{n}$\\
OMF & $\mathbf{C}_{\mathbf{x}}=\lambda_\mathbf{x}\mathbf{I}_X$, $\mathbf{C}_{\mathbf{n}}=\lambda_\mathbf{n}\mathbf{I}_Y$, $\lambda_\mathbf{x}\ll\lambda_\mathbf{n}$\\
\bottomrule
\end{tabular}
\label{tab:special-cases}
\end{table}

\subsubsection{Ordinary Least Squares}

In \gls{ols}, also simply known as \gls{ls}, or linear regression, the entries of $\mathbf{x}$ and $\mathbf{n}$ are uncorrelated and have equal variance, i.e., $\mathbf{C}_{\mathbf{x}}=\lambda_\mathbf{x}\mathbf{I}_X$ and $\mathbf{C}_{\mathbf{n}}=\lambda_\mathbf{n}\mathbf{I}_Y$, and the variance on the a priori information on $\mathbf{x}$ is much higher than on $\mathbf{n}$, i.e., $\lambda_\mathbf{x}\gg\lambda_\mathbf{n}$.
Thus, since $\lambda=\lambda_\mathbf{n}/\lambda_\mathbf{x}\ll 1$, \eqref{eq:RLS1} and \eqref{eq:RLS2} further simplify as
\begin{align}
\hat{\mathbf{x}}_{\text{OLS},1}
&=\left(\mathbf{H}^H\mathbf{H}\right)^{-1}\mathbf{H}^H\mathbf{y},\\
\hat{\mathbf{x}}_{\text{OLS},2}
&=\mathbf{H}^H\left(\mathbf{H}\mathbf{H}^H\right)^{-1}\mathbf{y},
\end{align}
respectively.
Interestingly, \gls{ols} is a special case of both \gls{gls} and \gls{rls} as it includes both their assumptions on $\mathbf{C}_{\mathbf{x}}$ and $\mathbf{C}_{\mathbf{n}}$.
As discussed for the \gls{gls}, since only one matrix among $\mathbf{H}^H\mathbf{H}$ and $\mathbf{H}\mathbf{H}^H$ may be invertible depending on the dimensions of $\mathbf{H}$, only one among $\hat{\mathbf{x}}_{\text{OLS},1}$ and $\hat{\mathbf{x}}_{\text{OLS},2}$ may exist, unless $\mathbf{H}$ is a square matrix.
Remarkably, the expressions of the \gls{ols} are widely employed in \gls{mimo} wireless communications to implement \gls{zfbf} and the \gls{zf} receiver \cite[Chapters~7,~12]{cle13}.

\subsubsection{Ordinary Matched Filtering}

In \gls{omf}, also simply known as \gls{mf}, the entries of $\mathbf{x}$ and $\mathbf{n}$ are uncorrelated and have equal variance, i.e., $\mathbf{C}_{\mathbf{x}}=\lambda_\mathbf{x}\mathbf{I}_X$ and $\mathbf{C}_{\mathbf{n}}=\lambda_\mathbf{n}\mathbf{I}_Y$, and the variance on the a priori information on $\mathbf{x}$ is much smaller than on $\mathbf{n}$, i.e., $\lambda_\mathbf{x}\ll\lambda_\mathbf{n}$.
Thus, since $\lambda=\lambda_\mathbf{n}/\lambda_\mathbf{x}\gg 1$, both \eqref{eq:RLS1} and \eqref{eq:RLS2} boil down to
\begin{equation}
\hat{\mathbf{x}}_{\text{OMF}}=\lambda^{-1}\mathbf{H}^H\mathbf{y}.\label{eq:sc8}
\end{equation}
Remarkably, \gls{omf} is a special case of both \gls{gmf} and \gls{rls} as it jointly accounts for their assumptions on $\mathbf{C}_{\mathbf{x}}$ and $\mathbf{C}_{\mathbf{n}}$.
In addition, it is utilized in \gls{mimo} wireless communications to realize \gls{mbf} and the \gls{mf} receiver \cite[Chapters~7,~12]{cle13}.

We summarize the five special cases of the \gls{lmmse} estimator for linear observation processes and their assumptions on $\mathbf{C}_{\mathbf{x}}$ and $\mathbf{C}_{\mathbf{n}}$ in Tab.~\ref{tab:special-cases}.

\begin{table}[t]
\centering
\caption{MiLAC output $\mathbf{v}_{2}$ for different $\mathbf{P}$ with $\mathbf{P}_{11}$ invertible.}
\begin{tabular}{@{}lcc@{}}
\toprule
 & $\mathbf{v}_{2}$ &
$\begin{bmatrix}
\mathbf{P}_{11} & \mathbf{P}_{12}\\
\mathbf{P}_{21} & \mathbf{P}_{22}
\end{bmatrix}$\\
\midrule
LMMSE  &
$\left(\mathbf{H}^H\mathbf{C}_{\mathbf{n}}^{-1}\mathbf{H}+\mathbf{C}_{\mathbf{x}}^{-1}\right)^{-1}\mathbf{H}^H\mathbf{C}_{\mathbf{n}}^{-1}\mathbf{u}$ &
$\begin{bmatrix}
\pm\mathbf{C}_{\mathbf{n}} & \mathbf{H}\\
\mathbf{H}^H & \mp\mathbf{C}_{\mathbf{x}}^{-1}
\end{bmatrix}$\\
\midrule
GLS &
$\left(\mathbf{H}^H\mathbf{C}_{\mathbf{n}}^{-1}\mathbf{H}\right)^{-1}\mathbf{H}^H\mathbf{C}_{\mathbf{n}}^{-1}\mathbf{u}$ &
$\begin{bmatrix}
\pm\mathbf{C}_{\mathbf{n}} & \mathbf{H}\\
\mathbf{H}^H & \mathbf{0}_{X}
\end{bmatrix}$\\
\midrule
GMF &
$\mathbf{C}_{\mathbf{x}}\mathbf{H}^H\mathbf{C}_{\mathbf{n}}^{-1}\mathbf{u}$ &
$\begin{bmatrix}
\pm\mathbf{C}_{\mathbf{n}} & \mathbf{0}_{Y\times X}\\
\mathbf{H}^H & \mp\mathbf{C}_{\mathbf{x}}^{-1}
\end{bmatrix}$\\
\midrule
RLS &
$\left(\mathbf{H}^H\mathbf{H}+\lambda\mathbf{I}_X\right)^{-1}\mathbf{H}^H\mathbf{u}$ &
$\begin{bmatrix}
\pm\mathbf{I}_Y & \mathbf{H}\\
\mathbf{H}^H & \mp\lambda\mathbf{I}_X
\end{bmatrix}$\\
\midrule
OLS &
$\left(\mathbf{H}^H\mathbf{H}\right)^{-1}\mathbf{H}^H\mathbf{u}$ &
$\begin{bmatrix}
\pm\mathbf{I}_Y & \mathbf{H}\\
\mathbf{H}^H & \mathbf{0}_X
\end{bmatrix}$\\
\midrule
OMF &
$\lambda^{-1}\mathbf{H}^H\mathbf{u}$ &
$\begin{bmatrix}
\pm\mathbf{I}_Y & \mathbf{0}_{Y\times X}\\
\mathbf{H}^H & \mp\lambda\mathbf{I}_X
\end{bmatrix}$\\
\bottomrule
\end{tabular}
\label{tab:v2-P11-inv}
\end{table}

\subsection{Computing the Special Cases of the LMMSE Estimator with a MiLAC}

The expressions of the five special cases of the \gls{lmmse} estimator can be efficiently returned on the output vector $\mathbf{v}_2$ of a \gls{milac} having $N=Y$ input ports and $M=X$ output ports with no input, i.e., $P=X+Y$ total ports.
This is achieved by exploiting \eqref{eq:v2-1} and \eqref{eq:v2-2}, as discussed for the \gls{lmmse} estimator in Part~I.
First, the tunable admittance components $\{Y_{i,k}\}_{i,k=1}^{P}$ are set according to \eqref{eq:Yik-component-P} such that the blocks of the matrix $\mathbf{P}$, namely $\mathbf{P}_{11}$, $\mathbf{P}_{12}$, $\mathbf{P}_{21}$, and $\mathbf{P}_{22}$, give the desired expression in the output $\mathbf{v}_2$.
Second, the input vector of the \gls{milac} is set as $\mathbf{u}=\mathbf{y}$.
In detail, the expressions of the special cases in \eqref{eq:sc1}-\eqref{eq:sc8} can be computed by setting the blocks of the matrix $\mathbf{P}$ as illustrated in Tabs.~\ref{tab:v2-P11-inv} and \ref{tab:v2-P22-inv}, depending on the invertibility of the blocks $\mathbf{P}_{11}$ and $\mathbf{P}_{22}$.
When the block $\mathbf{P}_{11}$ is invertible, $\mathbf{v}_2$ is given by \eqref{eq:v2-1}, as reported in Tab.~\ref{tab:v2-P11-inv}, while when the block $\mathbf{P}_{22}$ is invertible, $\mathbf{v}_2$ is given by \eqref{eq:v2-2}, as reported in Tab.~\ref{tab:v2-P22-inv}.
Note that \gls{gls} and \gls{ols} are computed via a matrix $\mathbf{P}$ whose blocks $\mathbf{P}_{11}$ and $\mathbf{P}_{22}$ are not both invertible.
Thus, for \gls{gls} and \gls{ols} the expressions in Tabs.~\ref{tab:v2-P11-inv} and \ref{tab:v2-P22-inv} are not equivalent in general, while they are equivalent for the \gls{lmmse} estimator and the other special cases.

\begin{remark}
The five special cases of the \gls{lmmse} estimator can be efficiently computed using a \gls{milac}, which offers significantly reduced computational complexity compared to a digital computer, in terms of the number of arithmetic real operations required.
This is because a \gls{milac} directly performs matrix-matrix multiplications and matrix inversion operations in the analog domain.
On the one hand, the computational complexity of computing the special cases with a \gls{milac} is driven by the number of operations required to compute \eqref{eq:Yik-component-P} for a given matrix $\mathbf{P}$.
Thus, the complexity required to compute \gls{gls}, \gls{rls}, and \gls{ols} with a \gls{milac} is the same as the complexity of computing the \gls{lmmse} estimator, i.e., $6XY$ real operations, as discussed in Part~I.
Besides, computing \eqref{eq:Yik-component-P} for the \gls{gmf} and \gls{omf} require approximately $2XY$ real operations for both cases $i\neq k$ and $i=k$, resulting in a total of $4XY$ real operations, where we assumed that the operations involving $\mathbf{C}_{\mathbf{x}}$ and $\mathbf{C}_{\mathbf{n}}$ are precomputed offline.
On the other hand, a digital computer calculates the \gls{gls} with the same complexity as the \gls{lmmse} estimator due to the two matrix-matrix products and matrix inversion, i.e., $8(XY^2+X^2Y+\min\{X^3,Y^3\}/3)$, as discussed in Part~I.
The \gls{rls} and \gls{ols} can be computed with reduced complexity as they require only one matrix-matrix product and a matrix inversion, namely $8\min\{X^2Y+X^3/3, XY^2+Y^3/3\}$.
Besides, the \gls{gmf} and \gls{omf} only require three and one matrix-vector products, respectively, resulting in a complexity of $8(X^2+XY+Y^2)$ and $8XY$ real operations, respectively.
\label{rem:special-cases}
\end{remark}

\begin{table}[t]
\centering
\caption{MiLAC output $\mathbf{v}_{2}$ for different $\mathbf{P}$ with $\mathbf{P}_{22}$ invertible.}
\begin{tabular}{@{}lcc@{}}
\toprule
 & $\mathbf{v}_{2}$ &
$\begin{bmatrix}
\mathbf{P}_{11} & \mathbf{P}_{12}\\
\mathbf{P}_{21} & \mathbf{P}_{22}
\end{bmatrix}$\\
\midrule
LMMSE &
$\mathbf{C}_{\mathbf{x}}\mathbf{H}^H\left(\mathbf{H}\mathbf{C}_{\mathbf{x}}\mathbf{H}^H+\mathbf{C}_{\mathbf{n}}\right)^{-1}\mathbf{u}$ &
$\begin{bmatrix}
\pm\mathbf{C}_{\mathbf{n}} & \mathbf{H}\\
\mathbf{H}^H & \mp\mathbf{C}_{\mathbf{x}}^{-1}
\end{bmatrix}$\\
\midrule
GLS &
$\mathbf{C}_{\mathbf{x}}\mathbf{H}^H\left(\mathbf{H}\mathbf{C}_{\mathbf{x}}\mathbf{H}^H\right)^{-1}\mathbf{u}$ &
$\begin{bmatrix}
\mathbf{0}_Y & \mathbf{H}\\
\mathbf{H}^H & \mp\mathbf{C}_{\mathbf{x}}^{-1}
\end{bmatrix}$\\
\midrule
GMF &
$\mathbf{C}_{\mathbf{x}}\mathbf{H}^H\mathbf{C}_{\mathbf{n}}^{-1}\mathbf{u}$ &
$\begin{bmatrix}
\pm\mathbf{C}_{\mathbf{n}} & \mathbf{0}_{Y\times X}\\
\mathbf{H}^H & \mp\mathbf{C}_{\mathbf{x}}^{-1}
\end{bmatrix}$\\
\midrule
RLS &
$\mathbf{H}^H\left(\mathbf{H}\mathbf{H}^H+\lambda\mathbf{I}_Y\right)^{-1}\mathbf{u}$ &
$\begin{bmatrix}
\pm\lambda\mathbf{I}_Y & \mathbf{H}\\
\mathbf{H}^H & \mp\mathbf{I}_X
\end{bmatrix}$\\
\midrule
OLS &
$\mathbf{H}^H\left(\mathbf{H}\mathbf{H}^H\right)^{-1}\mathbf{u}$ &
$\begin{bmatrix}
\mathbf{0}_Y & \mathbf{H}\\
\mathbf{H}^H & \mp\mathbf{I}_X
\end{bmatrix}$\\
\midrule
OMF &
$\lambda^{-1}\mathbf{H}^H\mathbf{u}$ &
$\begin{bmatrix}
\pm\lambda\mathbf{I}_Y & \mathbf{0}_{Y\times X}\\
\mathbf{H}^H & \mp\mathbf{I}_X
\end{bmatrix}$\\
\bottomrule
\end{tabular}
\label{tab:v2-P22-inv}
\end{table}

\section{A Novel Beamforming Strategy:\\MiLAC-Aided Beamforming}
\label{sec:gen-analog-bf}

In this section, we show how a \gls{milac} can be used to implement a novel beamforming technique, that we denote as \gls{milac}-aided beamforming, at the transmitter as well as receiver side.
\Gls{milac}-aided beamforming has the same flexibility as digital beamforming, while it offers circuit and computational complexity benefits by fully operating in the analog domain.
To illustrate \gls{milac}-aided beamforming, we consider a \gls{mimo} communication system between $N_T$ transmitting antennas and $N_R$ receiving antennas through the wireless channel $\mathbf{H}\in\mathbb{C}^{N_R\times N_T}$.
We denote as $\mathbf{x}\in\mathbb{C}^{N_T\times1}$ the transmitted signal and as $\mathbf{y}\in\mathbb{C}^{N_R\times1}$ the received signal such that $\mathbf{y}=\mathbf{H}\mathbf{x}+\mathbf{n}$, where $\mathbf{n}\in\mathbb{C}^{N_R\times1}$ is the \gls{awgn}\footnote{We consider here a general \gls{mimo} system, such as single-user \gls{mimo} or multi-user with one symbol per user.
Both single- and multi-user systems will be specifically studied in the following subsections.}.
We begin by showing how \gls{milac}-aided beamforming can be used to implement an arbitrary beamforming matrix.

\begin{figure}[t]
\centering
\includegraphics[width=0.34\textwidth]{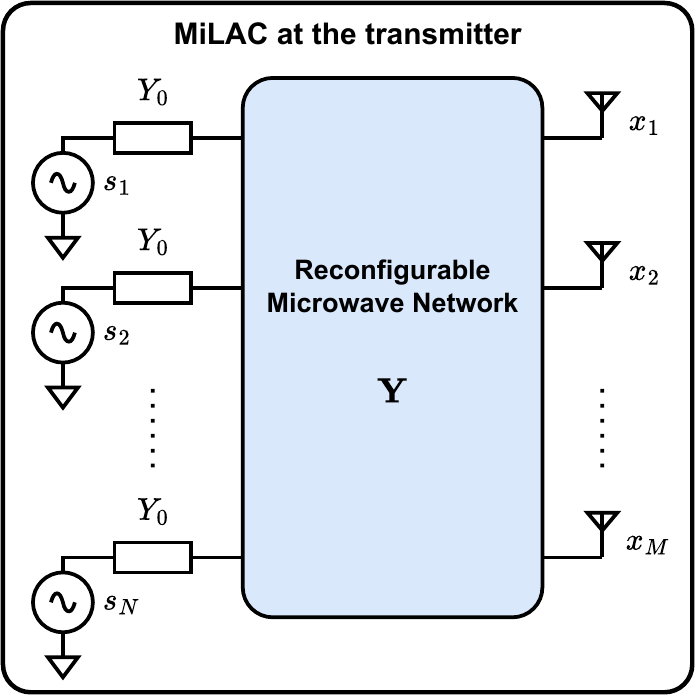}
\caption{MiLAC-aided beamforming at the transmitter, performing $\mathbf{x}=\mathbf{W}\mathbf{s}$ in the analog domain, where $\mathbf{s}$ is the symbol vector and $\mathbf{W}$ is an arbitrary precoding matrix.}
\label{fig:tx}
\end{figure}

\subsection{Arbitrary Beamforming}
\label{sec:arbitrary-bf}

Consider a transmitter that linearly precodes a vector of $N_S$ symbols $\mathbf{s}\in\mathbb{C}^{N_S\times1}$ to obtain the transmitted signal as $\mathbf{x}=\mathbf{W}\mathbf{s}$, where $\mathbf{W}\in\mathbb{C}^{N_T\times N_S}$ is an arbitrary precoding matrix.
Such a transmitter can be implemented purely in the analog domain by a \gls{milac} having $N=N_S$ ports receiving the symbol vector in input and $M=N_T$ ports delivering the output to antennas perfectly matched to the reference impedance $Z_0=Y_0^{-1}$, as represented in Fig.~\ref{fig:tx}.
Since the antennas are perfectly matched, their Thevenin equivalent circuit is an impedance $Z_0=Y_0^{-1}$ connected to ground \cite[Chapter~2.13]{bal15}.
Thus, the \gls{milac} in Fig.~\ref{fig:tx} behaves as when its last $N_T$ ports are connected to ground through an impedance $Z_0=Y_0^{-1}$, i.e., as analyzed in Section~\ref{sec:milac}.
Following the analysis in Section~\ref{sec:milac}, such a \gls{milac} returns $\mathbf{x}=\mathbf{W}\mathbf{s}$ on its last $N_T$ ports when the signal $\mathbf{s}$ is fed as input on the first $N_S$ ports and the tunable admittance components are set such that the matrix $\mathbf{P}$ is
\begin{equation}
\mathbf{P}=
\begin{bmatrix}
\mathbf{I}_{N_S} & \mathbf{0}_{N_S\times N_T}\\
-\mathbf{W} & \mathbf{I}_{N_T}
\end{bmatrix},\label{eq:P-beamforming}
\end{equation}
according to \eqref{eq:v2-1} and \eqref{eq:v2-2}.
To this end, the tunable admittance components are set based on \eqref{eq:Yik-component-P} where $\mathbf{P}$ is given by \eqref{eq:P-beamforming}, yielding
\begin{equation}
Y_{i,k}
=
\begin{cases}
Y_0\left[\mathbf{W}\right]_{i-N,k} & N_S<i,\;k\leq N_S\\
0 & \text{Otherwise}
\end{cases},\label{eq:Yik-arbitrary}
\end{equation}
for $i\neq k$, and
\begin{equation}
Y_{k,k}
=
\begin{cases}
-Y_0\sum_{n_T=1}^{N_T}\left[\mathbf{W}\right]_{n_T,k} & k\leq N_S\\
0 & k> N_S
\end{cases},\label{eq:Ykk-arbitrary}
\end{equation}
for $k=1,\ldots,N_S+N_T$.
From \eqref{eq:Yik-arbitrary} and \eqref{eq:Ykk-arbitrary}, we observe that the tunable admittance components of a \gls{milac} can be reconfigured in closed form to implement any arbitrary beamforming $\mathbf{W}$ matrix in the analog domain.
The computational complexity of this beamforming technique, denoted as \gls{milac}-aided beamforming, is given by the complexity of designing $\mathbf{W}$ (depending on the adopted precoding strategy) and computing \eqref{eq:Yik-arbitrary} and \eqref{eq:Ykk-arbitrary} (requiring in total $4N_SN_T$ real operations).
Note that these operations are executed at each coherence block, while no operation is required on a per-symbol basis since the matrix-vector product $\mathbf{W}\mathbf{s}$ is performed in the analog domain.
To adapt the beamforming matrix $\mathbf{W}$ to the wireless channel $\mathbf{H}$, the \gls{milac} is assumed to be implemented with tunable admittance components that can be reconfigured at every channel coherence time.

\begin{figure}[t]
\centering
\includegraphics[width=0.34\textwidth]{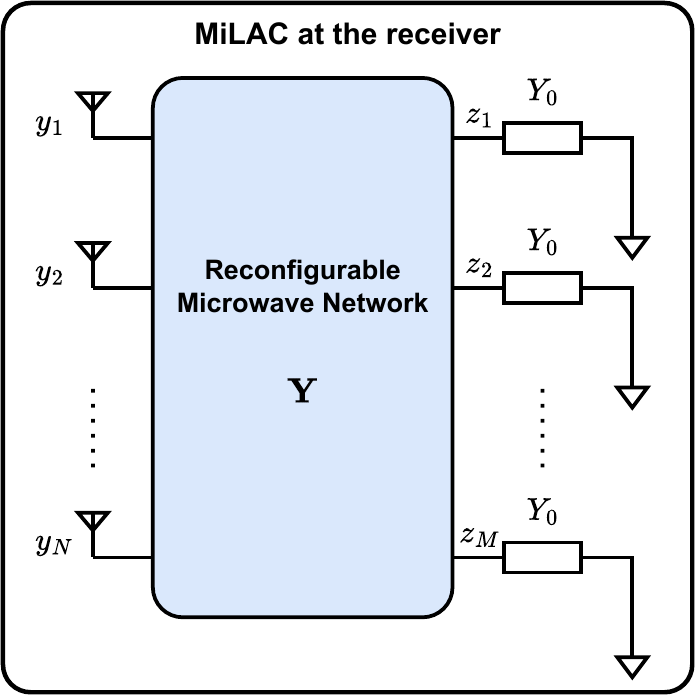}
\caption{MiLAC-aided beamforming at the receiver, performing $\mathbf{z}=\mathbf{G}\mathbf{y}$ in the analog domain, where $\mathbf{y}$ is the received signal and $\mathbf{G}$ is an arbitrary combining matrix.}
\label{fig:rx}
\end{figure}

As discussed on the transmitter side, \gls{milac}-aided beamforming can also be introduced on the receiver side.
Specifically, a \gls{milac} can be used to linearly combine the received signal $\mathbf{y}\in\mathbb{C}^{N_R\times1}$ to obtain the signal used for detection $\mathbf{z}\in\mathbb{C}^{N_S\times1}$ as $\mathbf{z}=\mathbf{G}\mathbf{y}$, where $\mathbf{G}\in\mathbb{C}^{N_S\times N_R}$ is an arbitrary combining matrix.
This can be obtained with \gls{milac} having $N=N_R$ ports acquiring the input from the receiving antennas and $M=N_S$ ports providing the output vector $\mathbf{z}$, as represented in Fig.~\ref{fig:rx}.
Note that in Fig.~\ref{fig:rx} we assume the antennas to be perfectly matched to the reference impedance $Z_0=Y_0^{-1}$, such that their Thevenin equivalent circuit is a voltage generator (inducing a voltage $y_n$ at port $n$, for $n=1,\ldots,N$) in series to an impedance $Z_0=Y_0^{-1}$ \cite[Chapter~2.13]{bal15}, i.e., as in the \gls{milac} analyzed in Section~\ref{sec:milac}.

Among the possible linear precoding and combining strategies, \gls{milac}-aided beamforming can efficiently perform \gls{r-zfbf}, \gls{zfbf}, and \gls{mbf} at the transmitter and \gls{mmse}, \gls{zf}, and \gls{mf} at the receiver by exploiting the capability of a \gls{milac} to compute the corresponding special case of the \gls{lmmse} estimator.
This leads to remarkable computational complexity benefits, as illustrated in the following.

\subsection{LMMSE-Inspired Beamforming at the Transmitter:\\R-ZFBF, ZFBF, and MBF Transmitters}
\label{sec:tx-analog-bf}

Assume that the previously considered \gls{mimo} system $\mathbf{y}=\mathbf{H}\mathbf{x}+\mathbf{n}$ is a multi-user system between an $N_T$-antenna transmitter and $N_R$ single-antenna users, with $N_R\leq N_T$.
The signal received at the $N_R$ users $\mathbf{y}$ is used to detect the symbols $\mathbf{s}\in\mathbb{C}^{N_S\times1}$, with $N_S=N_R$, and the transmitted signal is given by $\mathbf{x}=\mathbf{W}\mathbf{s}$, where $\mathbf{W}\in\mathbb{C}^{N_T\times N_R}$ is the precoding matrix implemented by a \gls{milac}, as in Fig.~\ref{fig:tx}.
By exploiting the operations included in Tab.~\ref{tab:v2-P22-inv}, it is possible to reconfigure the microwave network of the \gls{milac} to efficiently execute three popular linear transmitters \cite[Chapter~12]{cle13}.
First, a \gls{milac} computing \gls{rls} can realize \gls{r-zfbf}, with precoding matrix $\mathbf{W}_{\text{R-ZFBF}}=\mathbf{H}^H(\mathbf{H}\mathbf{H}^H+\lambda\mathbf{I}_{N_R})^{-1}$, where $\lambda$ is an arbitrary regularization factor.
Second, \gls{zfbf} can be implemented by a \gls{milac} computing \gls{ols}, whose precoding matrix is $\mathbf{W}_{\text{ZFBF}}=\mathbf{H}^H(\mathbf{H}\mathbf{H}^H)^{-1}$.
Third, a \gls{milac} computing \gls{omf} can realize \gls{mbf}, with precoding matrix given by $\mathbf{W}_{\text{MBF}}=\lambda^{-1}\mathbf{H}^H$.
These three transmitters are obtained by reconfiguring the tunable admittance components of the \gls{milac} according to the channel $\mathbf{H}$ on a per coherence block basis, and do not require any operation on a per-symbol basis.
In addition, they require ultra-low computational complexity on a per coherence block basis since the matrix $\mathbf{P}$ of the \gls{milac} is readily given as a function of the wireless channel $\mathbf{H}$.
Hence, the only operations required are to compute the admittance values $\{Y_{i,k}\}_{i,k=1}^{P}$ from $\mathbf{P}$.

\begin{remark}
According to Remark~\ref{rem:special-cases}, the \gls{r-zfbf} and \gls{zfbf} transmitters can be achieved with only $6N_TN_R$ real operations, while the \gls{mbf} transmitter with only $4N_TN_R$ real operations, which are necessary to set the tunable admittance components of the \gls{milac}.
Conversely, the computational complexity of digitally calculating the \gls{r-zfbf} and \gls{zfbf} transmitters is $8(N_TN_R^2+N_R^3/3)$ real operations, while for the \gls{mbf} transmitter is $8N_TN_R$.
This confirms the significant benefits of \gls{milac}-aided beamforming in terms of computational complexity, especially in implementing the \gls{r-zfbf} and \gls{zfbf} transmitters.
\label{rem:tx}
\end{remark}

\subsection{LMMSE-Inspired Beamforming at the Receiver:\\MMSE, ZF, and MF Receivers}
\label{sec:rx-analog-bf}

Assume that the \gls{mimo} system $\mathbf{y}=\mathbf{H}\mathbf{x}+\mathbf{n}$ is a single-user system between an $N_T$-antenna transmitter and a $N_R$-antenna receiver, with $N_R\geq N_T$.
The transmitted signal $\mathbf{x}$ contains the symbols to be detected at the receiver, i.e., $\mathbf{x}=\mathbf{s}$, and the receiver uses the signal $\mathbf{z}=\mathbf{G}\mathbf{y}\in\mathbb{C}^{N_T\times 1}$ to accurately detect the symbols, where $\mathbf{G}\in\mathbb{C}^{N_T\times N_R}$ is the combining matrix imposed by the \gls{milac}, as in Fig.~\ref{fig:rx}.
By using the operations in Tab.~\ref{tab:v2-P11-inv}, it is possible to reconfigure the microwave network of the \gls{milac} to efficiently implement three widely used receivers \cite[Chapter~7]{cle13}.
First, a \gls{milac} computing \gls{rls} can realize the \gls{mmse} receiver with combining matrix $\mathbf{G}_{\text{MMSE}}=(\mathbf{H}^H\mathbf{H}+\lambda\mathbf{I}_{N_T})^{-1}\mathbf{H}^H$.
If the covariance matrix of the transmitted signal $\mathbf{x}$ is $\mathbf{C}_{\mathbf{x}}=P_T/N_T\mathbf{I}_{N_T}$, where $P_T$ is the transmit power, and the covariance matrix of the noise $\mathbf{n}$ is $\mathbf{C}_{\mathbf{n}}=\sigma_{\mathbf{n}}^2\mathbf{I}_{N_R}$, where $\sigma_{\mathbf{n}}^2$ is the noise power, the optimal regularization factor $\lambda$ is given by $\lambda=N_T\sigma_{\mathbf{n}}^2/P_T$ \cite[Chapter~7]{cle13}.
Second, the \gls{zf} receiver can be realized by a \gls{milac} computing \gls{ols}, whose combining matrix is $\mathbf{G}_{\text{ZF}}=(\mathbf{H}^H\mathbf{H})^{-1}\mathbf{H}^H$.
Third, a \gls{milac} computing \gls{omf} can implement the \gls{mf} receiver, with combining matrix $\mathbf{G}_{\text{MF}}=\lambda^{-1}\mathbf{H}^H$, where the optimal $\lambda$ is $\lambda=N_T\sigma_{\mathbf{n}}^2/P_T$ if $\mathbf{C}_{\mathbf{x}}=P_T/N_T\mathbf{I}_{N_T}$ and $\mathbf{C}_{\mathbf{n}}=\sigma_{\mathbf{n}}^2\mathbf{I}_{N_R}$ \cite[Chapter~7]{cle13}.
These three receivers are realized by solely reconfiguring the \gls{milac} on a per coherence block basis, while no operation is required on a per-symbol basis.

\begin{remark}
Similarly to Remark~\ref{rem:tx}, the \gls{mmse} and \gls{zf} receivers can be obtained through a \gls{milac} with only $6N_TN_R$ real operations and the \gls{mf} receiver with only $4N_TN_R$ real operations, which are required to set the tunable admittance components.
In contrast, the computational complexity of digitally calculating the \gls{mmse} and \gls{zf} receivers is $8(N_T^2N_R+N_T^3/3)$ real operations, and for the \gls{mf} receiver is $8N_TN_R$, showing important benefits in using \gls{milac}-aided beamforming to implement the \gls{mmse} and \gls{zf} receivers.
\label{rem:rx}
\end{remark}

\begin{figure*}[t]
\centering
\includegraphics[width=0.98\textwidth]{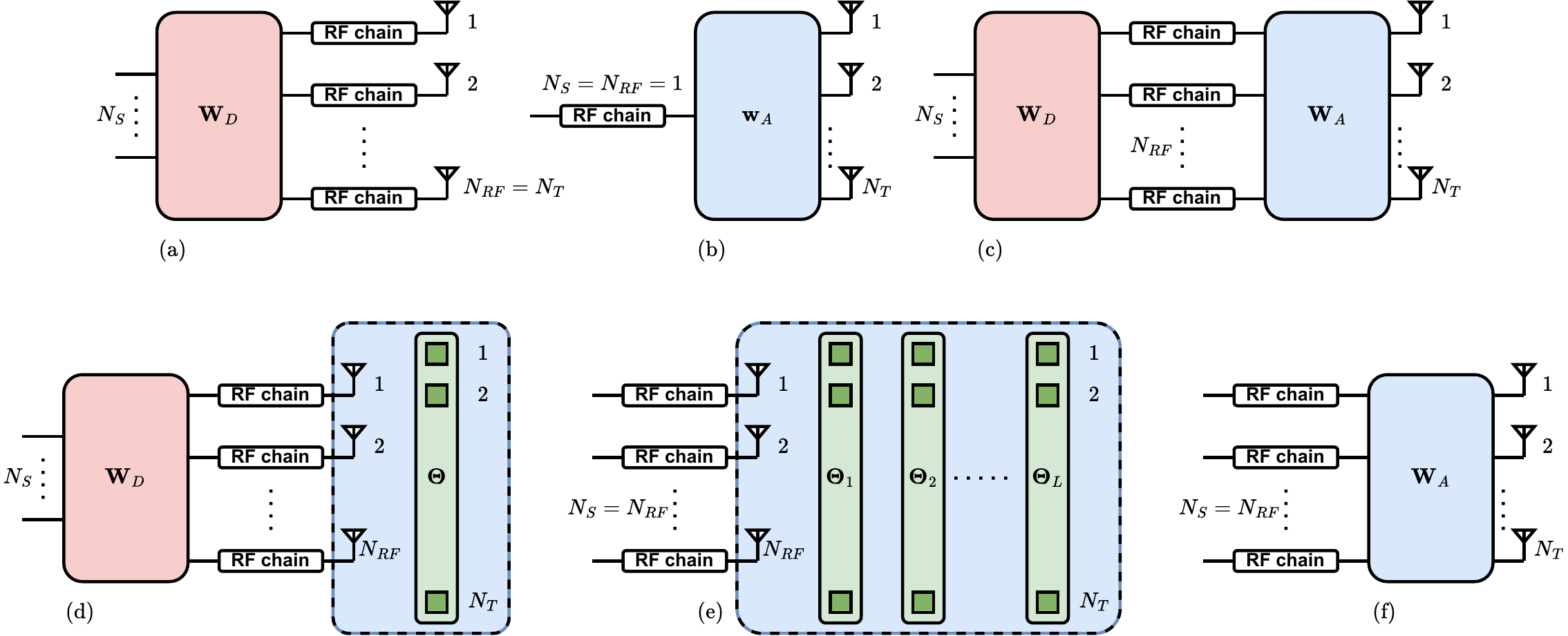}
\caption{(a) Digital, (b) analog, (c) hybrid, (d) RIS-aided, (e) SIM-aided, and (f) \gls{milac}-aided beamforming at the transmitter.}
\label{fig:comparison}
\end{figure*}

\begin{table*}[t]
\centering
\caption{Beamforming strategies comparison.}
\begin{tabular}{@{}lllllll@{}}
\toprule
Beamforming & Precoding matrix & Constraints &
\begin{tabular}{@{}l@{}}
Number of\\
RF chains
\end{tabular}
&
\begin{tabular}{@{}l@{}}
Resolution of\\
ADCs/DACs
\end{tabular}
&
\begin{tabular}{@{}l@{}}
Operations per\\
symbol basis
\end{tabular}
&
\begin{tabular}{@{}l@{}}
Operations per\\
coherence block
\end{tabular}\\
\midrule
Digital         & $\mathbf{W}_D$ & $\mathbf{W}_D\in\mathbb{C}^{N_T\times N_S}$ & $N_{RF}=N_T$ & High & Compute $\mathbf{W}_D\mathbf{s}$ & Optimize $\mathbf{W}_D$\\
\midrule
Analog          & $\mathbf{w}_A$ & $\mathbf{w}_A\in\mathbb{T}^{N_T\times 1}$ & $N_{RF}=N_S=1$ & Low & None & Optimize $\mathbf{w}_A$\\
\midrule
Hybrid          & $\mathbf{W}_A\mathbf{W}_D$ &
\begin{tabular}{@{}l@{}}
$\mathbf{W}_D\in\mathbb{C}^{N_{RF}\times N_S}$\\
$\mathbf{W}_A\in\mathbb{T}^{N_T\times N_{RF}}$
\end{tabular}
& $N_{RF}\geq N_S$ & High & Compute $\mathbf{W}_D\mathbf{s}$ & Optimize $\mathbf{W}_D,\mathbf{W}_A$\\
\midrule
RIS-aided       & $\boldsymbol{\Theta}\mathbf{H}_{\text{RIS}}\mathbf{W}_D$ &
\begin{tabular}{@{}l@{}}
$\mathbf{W}_D\in\mathbb{C}^{N_{RF}\times N_S}$\\
$\mathbf{H}_{\text{RIS}}\in\mathbb{C}^{N_T\times N_{RF}}$, fixed\\
$\boldsymbol{\Theta}\in\mathbb{C}^{N_T\times N_T},\boldsymbol{\Theta}^H\boldsymbol{\Theta}=\mathbf{I}_{N_T}$
\end{tabular}
& $N_{RF}\geq N_S$ & High & Compute $\mathbf{W}_D\mathbf{s}$ & Optimize $\mathbf{W}_D,\boldsymbol{\Theta}$\\
\midrule
SIM-aided       &
\begin{tabular}{@{}l@{}}
$\boldsymbol{\Theta}_{L}\mathbf{H}_{\text{SIM},L}\cdots$\\
$\cdots\boldsymbol{\Theta}_{1}\mathbf{H}_{\text{SIM},1}$
\end{tabular}
&
\begin{tabular}{@{}l@{}}
$\mathbf{H}_{\text{SIM},1}\in\mathbb{C}^{N_T\times N_{RF}}$, fixed\\
$\mathbf{H}_{\text{SIM},\ell}\in\mathbb{C}^{N_T\times N_T}$, $\ell>2$, fixed\\
$\boldsymbol{\Theta}_{\ell}\in\mathbb{C}^{N_T\times N_T},\boldsymbol{\Theta}_{\ell}^H\boldsymbol{\Theta}_{\ell}=\mathbf{I}_{N_T}$
\end{tabular}
& $N_{RF}=N_S$ & Low & None & Optimize $\boldsymbol{\Theta}_{\ell}$\\
\midrule
MiLAC-aided & $\mathbf{W}_A$ & $\mathbf{W}_A\in\mathbb{C}^{N_T\times N_S}$ & $N_{RF}=N_S$ & Low & None & Optimize $Y_{i,k}$\\
\bottomrule
\end{tabular}
\label{tab:comparison}
\end{table*}

\subsection{Computing the DFT}
\label{sec:dft}

\Gls{milac}-aided beamforming is also beneficial when used to precode or combine a signal with a fixed beamforming matrix.
In this case, since the admittance components of the \gls{milac} are optimized offline, no operation is required online with \gls{milac}-aided beamforming, while a matrix-vector product is needed on a per-symbol basis in the case of digital beamforming.
Combining the received signal $\mathbf{y}$ with a fixed matrix $\mathbf{G}$ is particularly useful for \gls{doa} estimation, where the received signal is typically multiplied by the \gls{dft} matrix \cite{an24b}.
This operation performed at the receiver can be formalized as $\mathbf{z}=\mathbf{G}\mathbf{y}$, where $\mathbf{G}\in\mathbb{C}^{N_R\times N_R}$ is the \gls{dft} matrix, given by
\begin{equation}
\left[\mathbf{G}\right]_{i,k}=\frac{1}{\sqrt{N_R}}e^{-j2\pi\frac{(i-1)(k-1)}{N_R}},\label{eq:dft}
\end{equation}
for $i,k=1,\ldots,N_R$.
This operation can be implemented purely in the analog domain by a \gls{milac} having $N=N_R$ ports acquiring the input $\mathbf{y}$ from the receiving antennas and $M=N_R$ ports returning the output, as represented in Fig.~\ref{fig:rx}.
To perform such a \gls{dft} operation, the admittance components of the \gls{milac} need to be set offline following \eqref{eq:Yik-component-P} where the matrix $\mathbf{P}$ given by 
\begin{equation}
\mathbf{P}=
\begin{bmatrix}
\mathbf{I}_{N_R} & \mathbf{0}_{N_R\times N_R}\\
-\mathbf{G} & \mathbf{I}_{N_R}
\end{bmatrix},\label{eq:P-dft}
\end{equation}
with $\mathbf{G}$ being the \gls{dft} matrix in \eqref{eq:dft}.
Thus, by substituting \eqref{eq:dft} and \eqref{eq:P-dft} into \eqref{eq:Yik-component-P}, we obtain that the admittance components $\{Y_{i,k}\}_{i,k=1}^{2N_R}$ are set as
\begin{equation}
Y_{i,k}
=
\begin{cases}
\frac{Y_0}{\sqrt{N_R}}e^{-j2\pi\frac{(i-N-1)(k-1)}{N_R}} & N_R<i,\;k\leq N_R\\
0 & \text{Otherwise}
\end{cases},\label{eq:Yik-dft}
\end{equation}
for $i\neq k$, and
\begin{equation}
Y_{k,k}
=
\begin{cases}
-Y_0\sqrt{N_R} & k=1\\
0 & k\neq 1
\end{cases},\label{eq:Ykk-dft}
\end{equation}
for $k=1,\ldots,2N_R$.

\begin{remark}
By employing a \gls{milac} with fixed admittance components set as in \eqref{eq:Yik-dft} and \eqref{eq:Ykk-dft}, we can perform the \gls{dft} instantly in the analog domain requiring no digital operations.
Conversely, the \gls{dft} requires approximately $34/9N_R\log_2(N_R)$ real operations to be digitally computed \cite{joh07}, which could become prohibitive as the number of antennas $N_R$ grows extremely large.
Note that we assume a one-dimensional \gls{dft} for simplicity, while a similar discussion also holds to perform a two-dimensional \gls{dft} in the analog domain, as considered in \cite{an24b}.
\label{rem:dft}
\end{remark}

\section{Comparison with Existing Beamforming Strategies}
\label{sec:comparison}

We have discussed how a \gls{milac} can enable \gls{milac}-aided beamforming, a new beamforming strategy purely in the analog domain having the same flexibility as digital beamforming.
In this section, we highlight the benefits of \gls{milac}-aided beamforming over the state-of-the-art beamforming strategies, showing that it requires fewer \gls{rf} chains, lower-resolution \glspl{adc}/\glspl{dac}, and reduced computational complexity.
To this end, we consider a transmitter with $N_T$ antennas and $N_{RF}$ \gls{rf} chains transmitting $N_S$ symbols, where we have $N_S\leq N_{RF}\leq N_T$\footnote{Several strategies have been proposed to perform symbol modulation in the analog domain, i.e., through \gls{espar} \cite{kaw05,han13,zha23}, \gls{lma} \cite{mul14,sed16}, and \gls{ris} \cite{tan20a,tan20b}, which can enable multi-stream transmissions ($N_S>1$) with $N_{RF}=1$ \gls{rf} chain.
However, these strategies operate symbol modulation on a per-symbol basis, and are fundamentally different from beamforming strategies acting per coherence block.
}.
We denote as $\mathbf{s}\in\mathbb{C}^{N_S\times1}$ the symbol vector and as $\mathbf{x}\in\mathbb{C}^{N_T\times1}$ the transmitted signal.
The following discussion for a transmitter readily applies also for a receiver.

\subsection{Digital Beamforming}

In digital beamforming, the transmitted signal is given by $\mathbf{x}=\mathbf{W}_D\mathbf{s}$, where $\mathbf{W}_D\in\mathbb{C}^{N_T\times N_S}$ is the digital precoder, subject only to the power constraint.
To allow maximum flexibility, there are $N_{RF}=N_T$ \gls{rf} chains equipped high-resolution \glspl{adc}/\glspl{dac}, each carrying an entry of the transmitted signal $\mathbf{x}$ to the corresponding antenna, as in Fig.~\ref{fig:comparison}(a).
In addition, the computational complexity is driven by the matrix-vector product $\mathbf{W}_D\mathbf{s}$ executed on a per-symbol basis and by the optimization of $\mathbf{W}_D$ at each coherence block based on the channel realization, whose complexity depends on the specific digital precoding strategy.

\subsection{Analog Beamforming}

In analog beamforming \cite{sun14}, only one symbol $s\in\mathbb{C}$ is transmitted ($N_S=1$), and the transmitted signal is given by $\mathbf{x}=\mathbf{w}_As$, where $\mathbf{w}_A\in\mathbb{C}^{N_T\times 1}$ is the analog precoder, as in Fig.~\ref{fig:comparison}(b).
Since the symbol $s$ reaches the $N_T$ antennas by passing through $N_T$ phase shifters, the analog precoder is subject to $\mathbf{w}_A\in\mathbb{T}^{N_T\times 1}$, where $\mathbb{T}=\{c\in\mathbb{C}:\vert c\vert=1\}$, due to the unit-modulus constraint imposed by the phase shifters.
In this case, a single \gls{rf} chain is sufficient to carry the symbol $s$, which can be equipped with low-resolution \glspl{adc}/\glspl{dac} since the symbol $s$ lies in a constellation with finite cardinality.
The computational complexity of analog beamforming is given by the complexity of optimizing $\mathbf{w}_A$ at every coherence block as a function of the channel realization.
Note that no operation takes place on a per-symbol basis since the product $\mathbf{w}_As$ is performed in the analog domain as the signal flows through the phase shifters connected to all antennas.

\subsection{Hybrid Beamforming}

In hybrid beamforming \cite{sun14,aya14,soh16,mol17,ahm18}, the transmitted signal is $\mathbf{x}=\mathbf{W}_A\mathbf{W}_D\mathbf{s}$, where $\mathbf{W}_A\in\mathbb{C}^{N_T\times N_{RF}}$ is the analog precoder, subject to specific constraints depending on the architecture of phase shifters and switches, and $\mathbf{W}_D\in\mathbb{C}^{N_{RF}\times N_S}$ is the digital precoder, subject only to the power constraint, as in Fig.~\ref{fig:comparison}(c).
Thus, the effective precoding matrix is given by $\mathbf{W}=\mathbf{W}_A\mathbf{W}_D$.
Numerous hybrid beamforming architectures have been proposed, each with different constraints on the matrix $\mathbf{W}_A$.
For example, in the fully-connected architecture, each \gls{rf} chain is connected to all the antennas through a phase shifter, and each entry of $\mathbf{W}_A$ is subject to the unit-modulus constraint, i.e., $\mathbf{W}_A\in\mathbb{T}^{N_T\times N_{RF}}$. 
The computational complexity of hybrid beamforming is given by the complexity of computing the matrix-vector product $\mathbf{W}_D\mathbf{s}$ on a per-symbol basis, and optimizing $\mathbf{W}_A$ and $\mathbf{W}_D$ at each coherence block.
This is because the digitally precoded signal $\mathbf{W}_D\mathbf{s}$ is multiplied by $\mathbf{W}_A$ in the analog domain.
Although the network of phase shifters in hybrid beamforming can operate a matrix-vector product in the analog domain, it cannot compute non-linear operations as a \gls{milac} made of admittance components.

\subsection{RIS-Aided Beamforming}

As a related analog-domain beamforming strategy, RIS-aided beamforming emerged by exploiting a \gls{ris} deployed in close proximity with an active transmitting device \cite{jam21,you22,hua23,mis24}, as in Fig.~\ref{fig:comparison}(d).
The \gls{ris} can work in either reflective or transmissive mode, without changing the system model.
Denoting as $\mathbf{x}^\prime=\mathbf{W}_D\mathbf{s}$ the digitally precoded signal transmitted by the active device, the received signal is given by $\mathbf{y}=\mathbf{H}_{\text{EFF}}\mathbf{x}^\prime$, where $\mathbf{H}_{\text{EFF}}\in\mathbb{C}^{N_R\times N_{RF}}$ is the RIS-aided channel, given by $\mathbf{H}_{\text{EFF}}=\mathbf{H}\boldsymbol{\Theta}\mathbf{H}_{\text{RIS}}$, with $\mathbf{H}\in\mathbb{C}^{N_R\times N_T}$, $\boldsymbol{\Theta}\in\mathbb{C}^{N_T\times N_T}$, and $\mathbf{H}_{\text{RIS}}\in\mathbb{C}^{N_T\times N_{RF}}$ being the channel from the \gls{ris} to the receiver, the scattering matrix of the RIS, and the channel from the active transmitting device to the \gls{ris}, respectively.
Thus, effective precoding matrix is $\mathbf{W}=\boldsymbol{\Theta}\mathbf{H}_{\text{RIS}}\mathbf{W}_D$, which is optimized by reconfiguring $\boldsymbol{\Theta}$, subject to $\boldsymbol{\Theta}^H\boldsymbol{\Theta}=\mathbf{I}_{N_T}$ in the case of an ideal lossless RIS \cite{she20}, $\mathbf{H}_{\text{RIS}}$ is not reconfigurable, and $\mathbf{W}_D$, subject only to the power constraint.
As in hybrid beamforming, the matrix-vector product $\mathbf{W}_D\mathbf{s}$ needs to be computed on a per-symbol basis, while $\mathbf{W}_D$ and $\boldsymbol{\Theta}$ are optimized per coherence block.

\subsection{SIM-Aided Beamforming}

\gls{ris}-aided beamforming has been extended by stacking multiple transmissive \glspl{ris} at the transmitter, which is denoted in the literature as a \gls{sim} \cite{an23,an24a,ner24}, as in Fig.~\ref{fig:comparison}(e).
Given the additional flexibility provided by the multiple metasurfaces, \gls{sim}-aided beamforming does not include any digital precoding, and for this reason it only requires $N_{RF}=N_S$ \gls{rf} chains equipped with low-resolution \glspl{adc}/\glspl{dac} \cite{an23,an24a,ner24,an24b}.
Similar to \gls{ris}-aided beamforming, \gls{sim}-aided beamforming exploits the fact that the effective channel between the transmitter and receiver can be engineered by reconfiguring the scattering matrices of the stacked \glspl{ris}.
Assuming a \gls{sim} made of $L$ stacked \glspl{ris} each with $N_T$ elements, the effective channel between the active device transmitting the symbols $\mathbf{s}$ and the receiver is given by
\begin{equation}
\mathbf{H}_{\text{EFF}}=\mathbf{H}\boldsymbol{\Theta}_{L}\mathbf{H}_{\text{SIM},L}\cdots\boldsymbol{\Theta}_{2}\mathbf{H}_{\text{SIM},2}\boldsymbol{\Theta}_{1}\mathbf{H}_{\text{SIM},1},\label{eq:Hsim}
\end{equation}
where $\mathbf{H}\in\mathbb{C}^{N_R\times N_T}$, $\boldsymbol{\Theta}_{\ell}\in\mathbb{C}^{N_T\times N_T}$, and $\mathbf{H}_{\text{SIM},\ell}\in\mathbb{C}^{N_T\times N_T}$ are the channel from the $L$th \gls{ris} to the receiver, the scattering matrix of the $\ell$th RIS, and the channel from the $\ell-1$th to the $\ell$th \gls{ris} (where the $0$th \gls{ris} is the transmitting device and $\mathbf{H}_{\text{SIM},1}\in\mathbb{C}^{N_T\times N_{RF}}$), respectively.
Thus, the effective precoding matrix reads as
\begin{equation}
\mathbf{W}=\boldsymbol{\Theta}_{L}\mathbf{H}_{\text{SIM},L}\cdots\boldsymbol{\Theta}_{2}\mathbf{H}_{\text{SIM},2}\boldsymbol{\Theta}_{1}\mathbf{H}_{\text{SIM},1},\label{eq:Wsim}
\end{equation}
which is optimized per coherence block by reconfiguring the scattering matrices of the $L$ stacked RISs  $\boldsymbol{\Theta}_{\ell}$, for $\ell=1,\ldots,L$.

Since \gls{sim}-aided beamforming is entirely performed in the analog domain, it can be regarded as a low-complexity approach for implementing \gls{milac}-aided beamforming, with fewer tunable admittance components.
Due to the reduced number of tunable components, the flexibility of the beamforming matrix is consequently reduced, and it is constrained as in \eqref{eq:Wsim} and cannot be an arbitrary matrix.

\begin{figure}[t]
\centering
\includegraphics[width=0.42\textwidth]{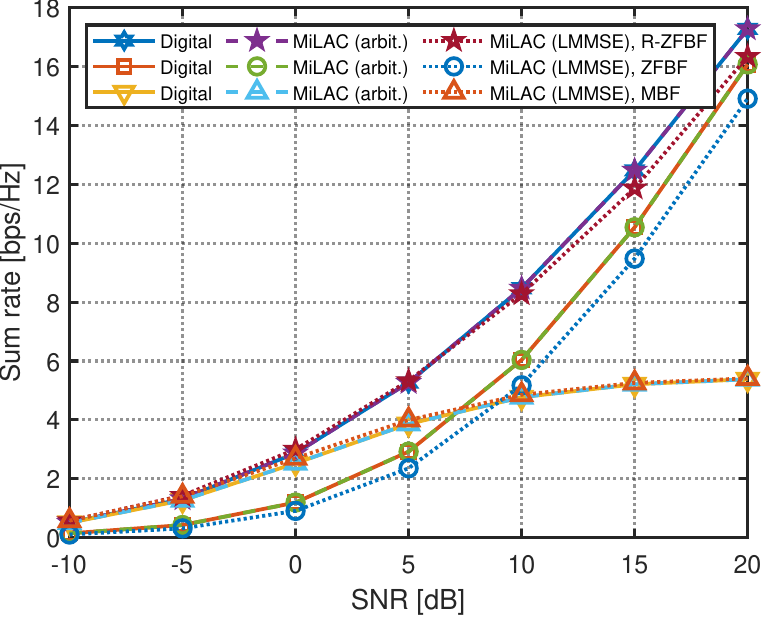}
\caption{Sum rate in a $4\times4$ multi-user MISO system, with digital and \gls{milac}-aided beamforming at the transmitter.}
\label{fig:tx-sr}
\end{figure}

\subsection{MiLAC-Aided Beamforming}

In \gls{milac}-aided beamforming, the symbols $\mathbf{s}$ are given in input to a \gls{milac}, and the transmitted signal $\mathbf{x}$ is returned on its output ports, as explained in Section~\ref{sec:gen-analog-bf}.
This novel beamforming strategy brings five unique benefits over state-of-the-art beamforming strategies.
Specifically, it offers the maximum flexibility with minimum hardware complexity (in terms of the number of \gls{rf} chains and resolution of \glspl{adc}/\glspl{dac}) and minimum computational complexity (in terms of operations per-symbol basis and per coherence block).

\subsubsection{Maximum Flexibility}

\Gls{milac}-aided beamforming has the same flexibility as digital beamforming since it can apply any arbitrary beamforming matrix $\mathbf{W}$, as illustrated in Section~\ref{sec:gen-analog-bf}.
Given its full flexibility, \gls{milac}-aided beamforming can obtain the same performance as digital beamforming, outperforming other beamforming strategies that impose limiting constraints on $\mathbf{W}$, i.e., analog, hybrid, \gls{ris}-aided, and \gls{sim}-aided, as summarized in Tab.~\ref{tab:comparison}.

\subsubsection{Minimum Number of RF Chains}

In \gls{milac}-aided beamforming, the symbols are fed in input to the \gls{milac}, which computes the transmitted signal in the analog domain.
Thus, only $N_{RF}=N_S$ \gls{rf} chains are needed to achieve the full flexibility of digital beamforming, which is the minimum number of \gls{rf} chains requested by the reviewed beamforming strategies, as shown in Tab.~\ref{tab:comparison}.
Note that a fully-connected hybrid beamforming architecture requires $N_{RF}=2N_S$ \gls{rf} chains to realize any digital beamforming matrix \cite{soh16}, while the other beamforming strategies are not proven to achieve full flexibility with any $N_{RF}<N_T$.

\subsubsection{Low-resolution ADCs/DACs}

Since \gls{milac}-aided beamforming processes the symbols purely in the analog domain, each \gls{rf} chain carries an individual symbol, which lies in a constellation with finite cardinality.
For example, in the case of \gls{qpsk}, the symbols can be chosen from the constellation $\{+1+j,+1-j,-1-j,-1+j\}$, up to a scaling factor, which can be generated through \glspl{dac} having just 1-bit resolution for both in-phase and quadrature parts of the signal.
Thus, low-resolution \glspl{adc}/\glspl{dac} can be adopted without sacrificing performance, which are less expensive and less power-hungry than high-resolution \glspl{adc}/\glspl{dac}.
High-resolution \glspl{adc}/\glspl{dac} are instead necessary when beamforming is performed in the digital domain (entirely or in part), i.e., in digital beamforming, hybrid beamforming, and \gls{ris}-aided beamforming, as indicated in Tab.~\ref{tab:comparison}.

\begin{figure}[t]
\centering
\includegraphics[width=0.42\textwidth]{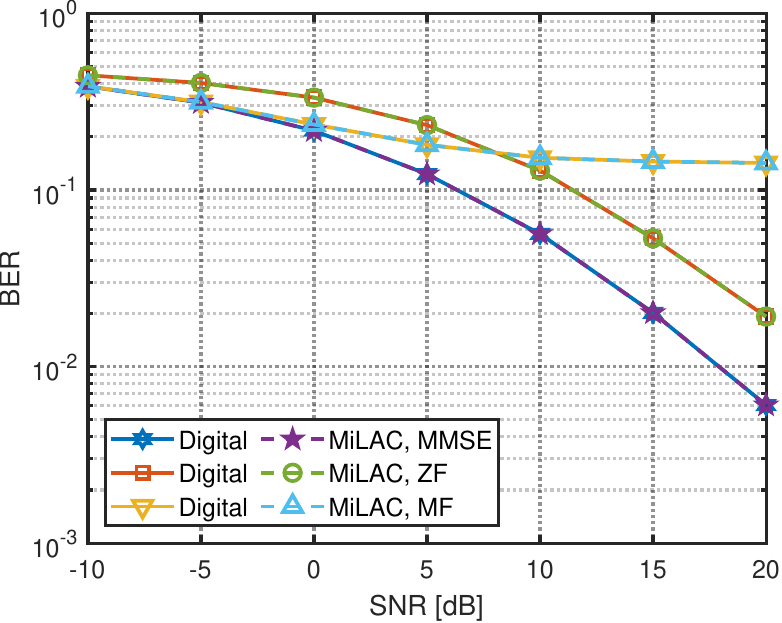}
\caption{BER in a $4\times4$ single-user MIMO system using QPSK, with digital and \gls{milac}-aided beamforming at the receiver.}
\label{fig:rx-ber}
\end{figure}

\subsubsection{No Computations on a Per-Symbol Basis}

If beamforming is performed in the digital domain (entirely or in part), the matrix-vector product $\mathbf{W}_D\mathbf{s}$ needs to be computed on a per-symbol basis, as shown in Tab.~\ref{tab:comparison}.
The complexity of this operation is $8N_{RF}N_S$ real operations, as discussed in the Appendix of Part~I, and can become prohibitive in massive/gigantic \gls{mimo} systems \cite{bjo16}.
\Gls{milac}-aided beamforming solves this issue since it does not require any computation on a per-symbol basis, as the beamforming is purely analog.
This reduces the beamforming computational complexity, especially when the coherence time is much longer than the symbol time \cite{bjo16}.

\subsubsection{Minimum Computations Per Coherence Block for RLS and OLS}

\Gls{milac}-aided beamforming offers further gains in terms of computational complexity by exploiting the capability of a \gls{milac} to efficiently compute \gls{rls} and \gls{ols} in the analog domain.
Specifically, these gains can be obtained when \gls{milac}-aided beamforming is used to compute the \gls{r-zfbf} and \gls{zfbf} transmitters or the \gls{mmse} and \gls{zf} receivers, as discussed in Section~\ref{sec:gen-analog-bf}.

\section{Numerical Results}
\label{sec:results}

We have proposed \gls{milac}-aided beamforming as a novel beamforming strategy that offers the same flexibility as digital beamforming, while presenting promising advantages in terms of circuit and computational complexity.
In this section, we evaluate the performance and computational complexity of \gls{milac}-aided beamforming.

\subsection{Performance Evaluation}
\label{sec:perf-eval}

Since \gls{milac}-aided beamforming can implement any arbitrary beamforming matrix, it is proven to perform as digital beamforming and outperform other beamforming strategies that are limited by constrained beamforming matrices.
The specific performance gains over the reviewed beamforming strategies match those of digital beamforming, which depend on the performance metric and scenario, such as whether the transmission is single- or multi-stream, and whether the system is single- or multi-user.
Thus, we analyze the cases where \gls{milac}-aided beamforming leverages the computing capabilities of a \gls{milac} to implement the \gls{lmmse}-inspired beamforming matrices, i.e., \gls{r-zfbf}, \gls{zfbf}, and \gls{mbf} at the transmitter, and the \gls{mmse}, \gls{zf}, and \gls{mf} at the receiver.

To evaluate the performance of \gls{r-zfbf}, \gls{zfbf}, and \gls{mbf} realized with digital and \gls{milac}-aided beamforming, we consider a multi-user \gls{miso} system with an $N_T$-antenna transmitter and $N_R$ single-antenna receivers, as presented in Section~\ref{sec:tx-analog-bf}.
Here, the transmitter precodes the $N_R$ symbols $\mathbf{s}\in\mathbb{C}^{N_R\times1}$ intended to the $N_R$ receivers through \gls{r-zfbf} or \gls{zfbf} to suppress inter-stream interference, or through \gls{mbf} to maximize the channel gains.
With digital beamforming, the \gls{r-zfbf} transmitter is realized by computing $\mathbf{F}_{\text{R-ZFBF}}=\mathbf{H}^H(\mathbf{H}\mathbf{H}^H+\lambda\mathbf{I}_{N_R})^{-1}$ and setting the precoding matrix $\mathbf{W}_{\text{R-ZFBF}}$ such that $[\mathbf{W}_{\text{R-ZFBF}}]_{:,n_R}=[\mathbf{F}_{\text{R-ZFBF}}]_{:,n_R}/\Vert[\mathbf{F}_{\text{R-ZFBF}}]_{:,n_R}\Vert_2$, for $n_R=1,\ldots,N_R$, assuming uniform power allocation to the users.
For large $N_R$ and if the \gls{snr} is the same at all receiving antennas, the optimal $\lambda$ is $\lambda=N_R\sigma_{\mathbf{n}}^2/P_T$, where $P_T$ is the transmit power, i.e., $\mathbf{s}$ has covariance matrix $\mathbf{C}_{\mathbf{s}}=P_T/N_R\mathbf{I}_{N_R}$, $\sigma_{\mathbf{n}}^2$ is the noise power, and we assumed the channel to have unit path gain \cite[Chapter~12]{cle13}.
Similarly, digital beamforming can realize \gls{zfbf} by computing $\mathbf{F}_{\text{ZFBF}}=\mathbf{H}^H(\mathbf{H}\mathbf{H}^H)^{-1}$ and setting $\mathbf{W}_{\text{ZFBF}}$ as $[\mathbf{W}_{\text{ZFBF}}]_{:,n_R}=[\mathbf{F}_{\text{ZFBF}}]_{:,n_R}/\Vert[\mathbf{F}_{\text{ZFBF}}]_{:,n_R}\Vert_2$, for $n_R=1,\ldots,N_R$, and \gls{mbf} by using $\mathbf{F}_{\text{MBF}}=\mathbf{H}^H$ and $\mathbf{W}_{\text{MBF}}$ such that $[\mathbf{W}_{\text{MBF}}]_{:,n_R}=[\mathbf{F}_{\text{MBF}}]_{:,n_R}/\Vert[\mathbf{F}_{\text{MBF}}]_{:,n_R}\Vert_2$, for $n_R=1,\ldots,N_R$.
When a \gls{milac} is used to perform \gls{r-zfbf}, \gls{zfbf}, and \gls{mbf}, we have two choices.
First, the \gls{milac} can be reconfigured to implement an arbitrary beamforming matrix, as discussed in Section~\ref{sec:arbitrary-bf}.
Thus, its tunable admittance components can be set according to \eqref{eq:Yik-arbitrary} and \eqref{eq:Ykk-arbitrary}, where $\mathbf{W}$ is the desired beamforming matrix $\mathbf{W}_{\text{R-ZFBF}}$, $\mathbf{W}_{\text{ZFBF}}$, or $\mathbf{W}_{\text{MBF}}$ with uniform power allocation, which is digitally computed.
Second, the \gls{milac} can directly compute the precoding matrices of these \gls{lmmse}-inspired beamforming techniques in the analog domain, as explained in Section~\ref{sec:tx-analog-bf}.
In this case, the columns of $\mathbf{W}_{\text{R-ZFBF}}$, $\mathbf{W}_{\text{ZFBF}}$, and  $\mathbf{W}_{\text{MBF}}$ cannot be individually normalized, and the resulting precoding matrices are given by $\mathbf{W}_{\text{R-ZFBF}}=\sqrt{N_R}\mathbf{F}_{\text{R-ZFBF}}/\Vert\mathbf{F}_{\text{R-ZFBF}}\Vert_F$, $\mathbf{W}_{\text{ZFBF}}=\sqrt{N_R}\mathbf{F}_{\text{ZFBF}}/\Vert\mathbf{F}_{\text{ZFBF}}\Vert_F$, and $\mathbf{W}_{\text{MBF}}=\sqrt{N_R}\mathbf{F}_{\text{MBF}}/\Vert\mathbf{F}_{\text{MBF}}\Vert_F$, where the normalization factor has been included to ensure that the transmit power is the same as with digital beamforming.

In Fig.~\ref{fig:tx-sr}, we report the sum rate versus the \gls{snr} $P_T/\sigma_{\mathbf{n}}^2$ achieved by digital and \gls{milac}-aided beamforming, where $N_T=N_R=4$ and the channels are \gls{iid} Rayleigh distributed with unit path gain.
We observe that \gls{r-zfbf} is the best precoder, which boils down to \gls{zfbf} at high \gls{snr} and to \gls{mbf} at low \gls{snr}, for both digital and \gls{milac}-aided beamforming, as expected from \gls{mimo} theory \cite[Chapter~12]{cle13}.
\gls{milac}-aided beamforming performs exactly as digital beamforming when the \gls{milac} is reconfigured based on the arbitrary beamforming matrix computed digitally, denoted as \textbf{MiLAC (arbit.)} in Fig.~\ref{fig:tx-sr}.
Besides, \gls{milac}-aided beamforming suffers a slight performance degradation at high \gls{snr} when the \gls{lmmse}-inspired beamforming matrix is computed in the analog domain, referred to as \textbf{MiLAC (LMMSE)} in Fig.~\ref{fig:tx-sr}.
This occurs because, in this case, \gls{milac}-aided beamforming cannot perform uniform power allocation, which is optimal at high \gls{snr}.
At the cost of this slight performance degradation, computing the \gls{lmmse}-inspired beamforming matrices in the analog domain substantially reduces the required computational complexity, as it will be analyzed in the following.

\begin{figure}[t]
\centering
\includegraphics[width=0.42\textwidth]{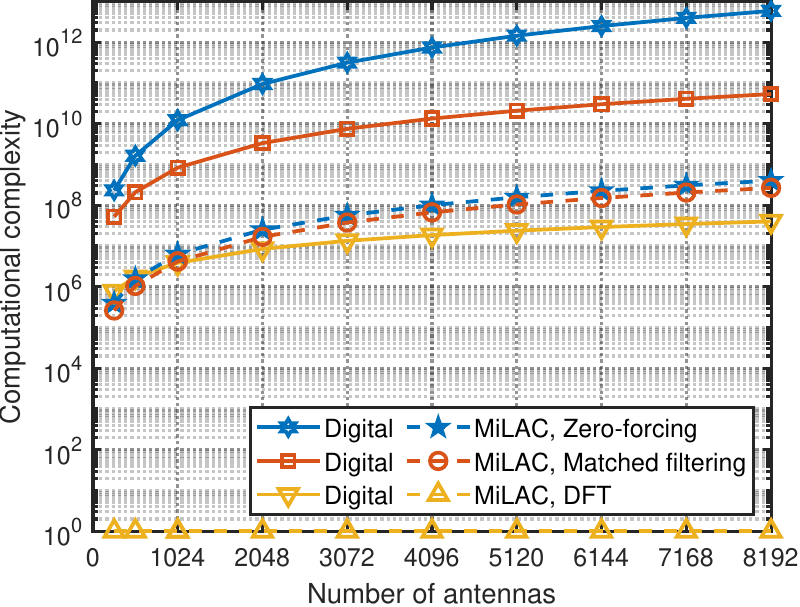}
\caption{Computational complexity per coherence block with length $\tau=100$ symbols of digital beamforming and \gls{milac}-aided beamforming}.
\label{fig:complexity-beamforming}
\end{figure}

To compare \gls{milac}-aided beamforming and digital beamforming at the receiver, we consider a single-user \gls{mimo} system with an $N_T$-antenna transmitter and an $N_R$-antenna receiver, where the receiver combines the received signal through \gls{mmse}, \gls{zf}, or \gls{mf}, as introduced in Section~\ref{sec:rx-analog-bf}.
In this case, the \gls{lmmse}-inspired combining matrices $\mathbf{G}_{\text{MMSE}}$, $\mathbf{G}_{\text{ZF}}$, and $\mathbf{G}_{\text{MF}}$ defined in Section~\ref{sec:rx-analog-bf} for \gls{milac}-aided beamforming are also the optimal expressions of the \gls{mmse}, \gls{zf}, and \gls{mf} receivers realized with digital beamforming.

In Fig.~\ref{fig:rx-ber}, we report the \gls{ber} versus the \gls{snr} $P_T/\sigma_{\mathbf{n}}^2$ obtained with digital and \gls{milac}-aided beamforming with $N_T=N_R=4$ and \gls{qpsk} modulation.
Since \gls{milac}-aided beamforming can implement the same combining matrices as digital beamforming, the two beamforming strategies achieve exactly the same \gls{ber}.

\begin{figure*}[t]
\centering
\subfigure[$\text{SNR}=0$~dB]{
\includegraphics[width=0.31\textwidth]{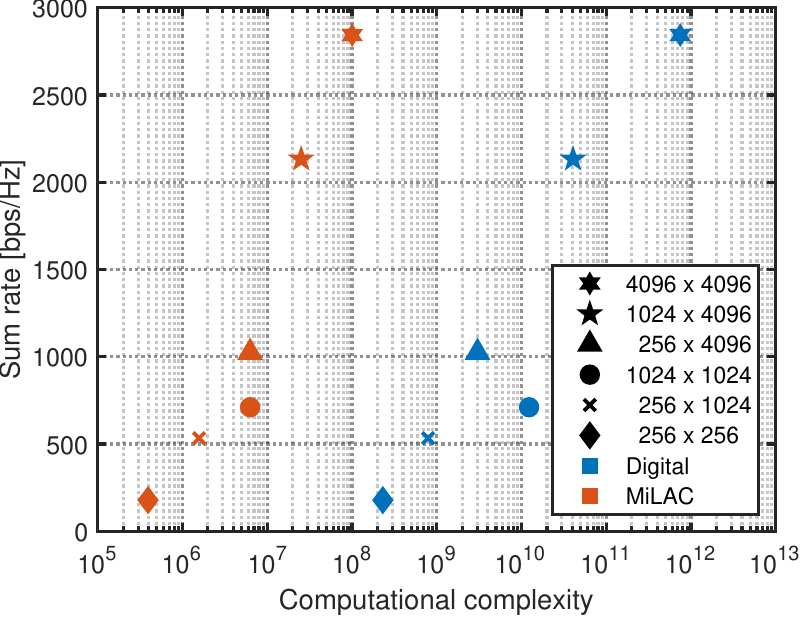}
\label{fig:perf-compl-0}}
\subfigure[$\text{SNR}=10$~dB]{
\includegraphics[width=0.31\textwidth]{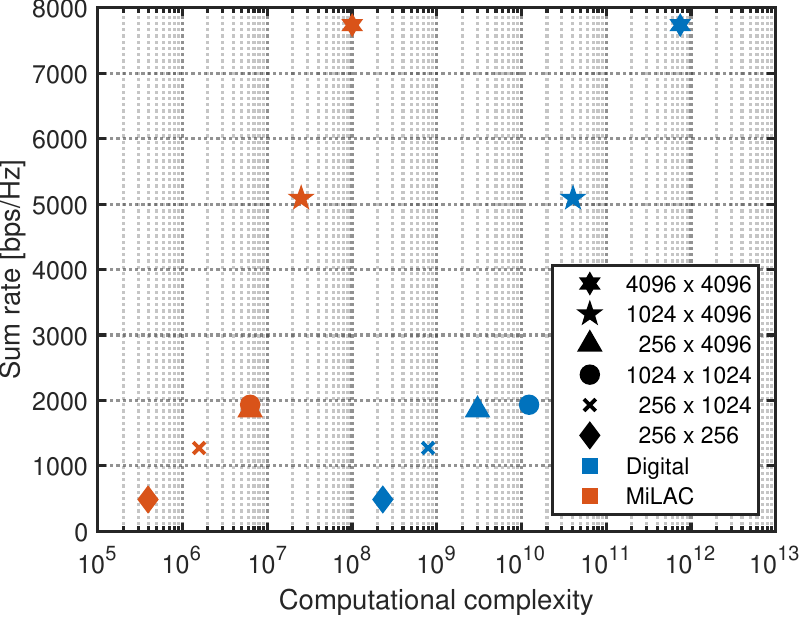}
\label{fig:perf-compl-10}}
\subfigure[$\text{SNR}=20$~dB]{
\includegraphics[width=0.31\textwidth]{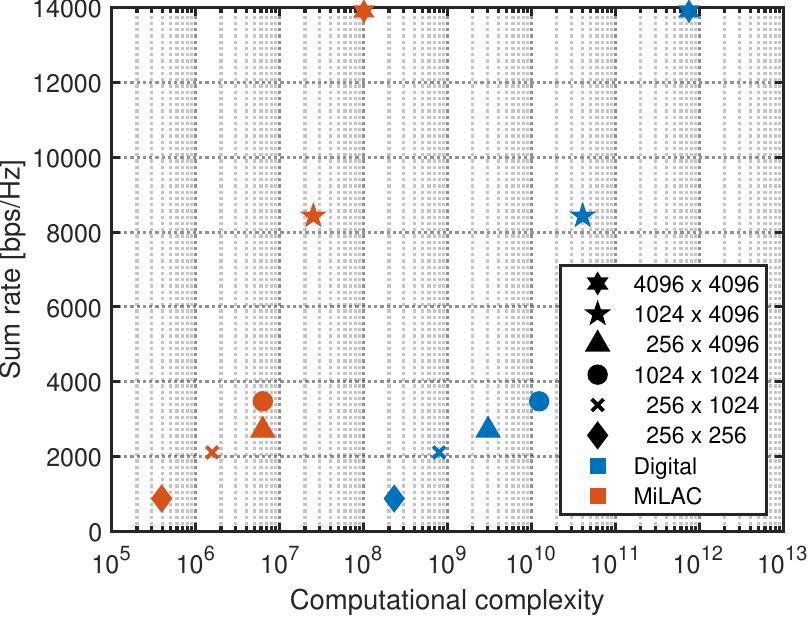}
\label{fig:perf-compl-20}}
\caption{Sum rate versus computational complexity of R-ZFBF performed in a $N_R\times N_T$ system with digital and MiLAC-aided beamforming, for three different levels of SNR.
Markers in blue correspond to digital beamforming and in red to MiLAC-aided beamforming.}
\label{fig:perf-compl}
\end{figure*}

\subsection{Computational Complexity Evaluation}

To quantify the benefits of \gls{milac}-aided beamforming in terms of computational complexity, we consider a \gls{mimo} system with  $N_T$ transmitting antennas and $N_R$ receiving antennas, with $N_T=N_R$, where \gls{milac}-aided beamforming is used at the transmitter to perform \gls{r-zfbf}, \gls{zfbf}, and \gls{mbf}, at the receiver for \gls{mmse}, \gls{zf}, and \gls{mf}, and for \gls{dft}, as discussed in Section~\ref{sec:tx-analog-bf}, \ref{sec:rx-analog-bf}, and \ref{sec:dft}.
In the considered system, the beamforming matrix is reconfigured per coherence block, where each coherence block includes $\tau$ symbol transmissions.
We define the computational complexity as the number of real operations per coherence block, as analyzed in the following.
\begin{itemize}
\item \textbf{Zero-forcing}. For \gls{r-zfbf} and \gls{zfbf} at the transmitter, or \gls{mmse} and \gls{zf} at the receiver, \gls{milac}-aided beamforming requires $6N_R^2$ operations per coherence block, as specified in Remarks~\ref{rem:tx} and \ref{rem:rx}, and no operation is required on a per-symbol basis.
Conversely, with digital beamforming, these strategies require $8(N_R^3+N_R^3/3)$ operations to design the beamforming matrix at each coherence block (see Remarks~\ref{rem:tx} and \ref{rem:rx}).
In addition, a matrix-vector product ($\mathbf{W}\mathbf{s}$ or $\mathbf{G}\mathbf{y}$) is performed on a per-symbol basis, requiring $8N_R^2$ real operations per symbol time and making the complexity per coherence block of digital beamforming $8(N_R^3+N_R^3/3)+8N_R^2\tau$.
\item \textbf{Matched filtering}. For \gls{mbf} at the transmitter, or \gls{mf} at the receiver, \gls{milac}-aided beamforming requires $4N_R^2$ operations per coherence block, as given in Remarks~\ref{rem:tx} and \ref{rem:rx}, with no operation on a per-symbol basis.
Besides, performing these strategies with digital beamforming requires $8N_R^2$ real operations per symbol time to perform the matrix-vector product $\mathbf{W}\mathbf{s}$ or $\mathbf{G}\mathbf{y}$, while no computation is needed to design the beamforming matrix, making the complexity per coherence block $8N_R^2\tau$.
\item \textbf{DFT}. Performing the \gls{dft} with \gls{milac}-aided beamforming requires no operation, since the \gls{milac} is not reconfigured at run-time (see Remark~\ref{rem:dft}).
Conversely, computing digitally the \gls{dft} requires $34/9N_R\log_2(N_R)$ operations per symbol time \cite{joh07}, giving $34/9N_R\log_2(N_R)\tau$ operations per coherence block.
\end{itemize}

In Fig.~\ref{fig:complexity-beamforming}, we show the computational complexity of digital and \gls{milac}-aided beamforming, as a function of the number of antennas $N_T=N_R$, by fixing $\tau=100$ symbols per coherence block.
For the cases \textbf{zero-forcing} and \textbf{matched filtering}, \gls{milac}-aided beamforming demonstrates up to $1.5\times10^4$ and $2.0\times10^2$ times lower computational complexity compared to digital beamforming, respectively, when the number of antennas is $N_R=8192$.
This significant reduction comes because of two reasons.
First, \gls{milac}-aided beamforming does not require the computation of the product $\mathbf{W}\mathbf{s}$, or $\mathbf{G}\mathbf{y}$, on a per-symbol basis, which has significant complexity given the high value of $\tau$ \cite{bjo16}.
Second, for \textbf{zero-forcing}, the complexity of \gls{milac}-aided beamforming scales with $\mathcal{O}(N_R^2)$ instead of $\mathcal{O}(N_R^3)$ since the expensive computation of the beamforming matrix is not required.
Besides, in the case of \textbf{DFT}, \gls{milac}-aided beamforming does not require any computation, saving up to $4.0\times10^7$ operations per coherence block when $N_R=8192$.
Given this significant reduction in computational complexity, we expect a proportional reduction in \gls{cpu} energy consumption and processing time.

\subsection{Relationship Between Performance and Computational Complexity}

We have shown that \gls{milac}-aided beamforming can achieve the same performance as digital beamforming in Figs.~\ref{fig:tx-sr} and \ref{fig:rx-ber}, and we have quantified its gain in terms of computational complexity in Fig.~\ref{fig:complexity-beamforming}.
We now investigate how the achieved performance and the required computational complexity are related in digital and \gls{milac}-aided beamforming.
To this end, we consider multi-user \gls{miso} system with an $N_T$-antenna transmitter serving $N_R$ single-antenna receivers through \gls{r-zfbf}.
For digital beamforming, we consider the \gls{r-zfbf} beamforming matrix to have uniform power allocation, as in Section~\ref{sec:perf-eval}.
For \gls{milac}-aided beamforming, we consider the \gls{lmmse}-inspired design discussed in Section~\ref{sec:tx-analog-bf}, where the computation of the \gls{r-zfbf} beamforming matrix is done in the analog domain.

In Fig.~\ref{fig:perf-compl}, we report the achieved sum rate and required computational complexity for 
\gls{milac}-aided and digital \gls{r-zfbf} beamforming in the case of different values of $N_R$ and $N_T$, where $N_R,N_T\in\{256,1024,4096\}$ and $N_R\leq N_T$, and \gls{snr}, where $\text{SNR}\in\{0,10,20\}$~dB.
We make the following three observations.
First, the sum rate increases with the number of transmit antennas $N_T$, the number of receivers $N_R$, and the \gls{snr}, for both \gls{milac}-aided and digital beamforming.
Besides, the computational complexity increases with the number of transmit antennas $N_T$ and receivers $N_R$, for both \gls{milac}-aided and digital beamforming, while is independent of the \gls{snr}.
Second, with a high number of transmit antennas, i.e., $N_T\geq256$, the performance achieved by a \gls{milac} computing the \gls{r-zfbf} matrix in the analog domain as per Section~\ref{sec:tx-analog-bf} is the same as with digital beamforming with uniform power allocation.
Thus, we can fully exploit the computational benefits of \gls{milac}-aided beamforming without sacrificing performance.
Third, \gls{milac}-aided beamforming offers the same performance as digital beamforming for all the considered configurations of $N_R$, $N_T$, and \gls{snr}, with significant gains in computational complexity.
Under a different perspective, \gls{milac}-aided beamforming can significantly improve the sum rate over digital beamforming given the same computational complexity.
For example, \gls{milac}-aided beamforming can operate \gls{r-zfbf} in a system with $N_R=N_T=4096$ requiring approximately the same complexity as digital \gls{r-zfbf} in a system with $N_R=N_T=256$.

\subsection{Performance Evaluation with Impairments}

So far, we have implicitly assumed perfect \gls{csi} and that the tunable admittance components of the \gls{milac} can be reconfigured with arbitrary precision.
We now compare \gls{milac}-aided beamforming with digital beamforming by relaxing these two assumptions, namely in the presence of noisy \gls{csi} and discrete-value admittance components in the \gls{milac}.

\begin{figure}[t]
\centering
\includegraphics[width=0.42\textwidth]{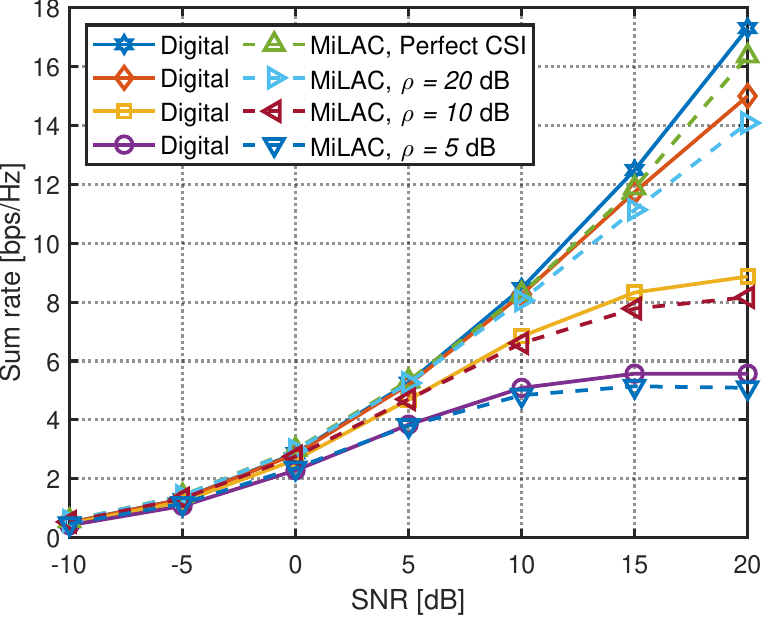}
\caption{Sum rate in a $4\times4$ multi-user MISO system, with noisy CSI.}
\label{fig:tx-sr-noisy-csi}
\end{figure}

Noisy \gls{csi} is modeled by adding \gls{awgn} to the perfect \gls{csi} $\mathbf{H}$, with different levels of \gls{snr} $\rho\in\{20,10,5\}$~dB.
In Fig.~\ref{fig:tx-sr-noisy-csi}, we report the sum rate achieved by \gls{r-zfbf} implemented with \gls{milac}-aided beamforming and digital beamforming optimized based on noisy \gls{csi}.
For \gls{milac}-aided beamforming, we consider that the MiLAC is reconfigured to operate \gls{lmmse}-inspired beamforming, as explained in Section~\ref{sec:tx-analog-bf}.
We observe that with $\rho=20$~dB the performance obtained by \gls{milac}-aided beamforming and digital beamforming is similar to the performance in the case of perfect \gls{csi}.
Besides, the performance degrades as $\rho$ decreases.
\Gls{milac}-aided beamforming approximately performs as digital beamforming for all the values of $\rho$, with a slight performance degradation at high \gls{snr}, as observed for perfect \gls{csi} in Fig.~\ref{fig:tx-sr}.

\begin{figure}[t]
\centering
\includegraphics[width=0.42\textwidth]{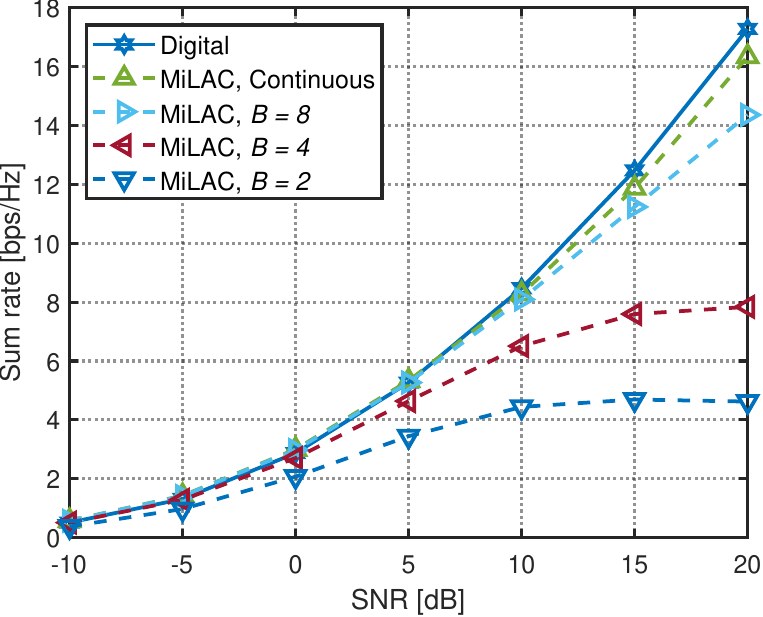}
\caption{Sum rate in a $4\times4$ multi-user MISO system, where the MiLAC has discrete-valued tunable admittance components.}
\label{fig:tx-sr-discrete}
\end{figure}

For practical implementation, it is convenient to design a \gls{milac} with tunable admittance components whose values are selected from a codebook of limited cardinality.
To this end, we notice that the optimal admittance components $Y_{i,k}$ with $i\neq k$ are distributed as the entries of the channel $\mathbf{H}$ when the \gls{milac} is used to perform \gls{r-zfbf}, following \eqref{eq:Yik-component-P} and Tab.~\ref{tab:v2-P22-inv}.
Thus, $Y_{i,k}$ with $i\neq k$ have real and imaginary parts Gaussian distributed in the case of \gls{iid} Rayleigh fading channels, which can be optimally quantized according to \cite{max60}.
In Fig.~\ref{fig:tx-sr-discrete}, we report the sum rate achieved by \gls{r-zfbf} implemented with \gls{milac}-aided beamforming where the admittance components $Y_{i,k}$ with $i\neq k$ are discretized with $B$ bits of resolution, where $B\in\{8,4,2\}$.
For the tunable components $Y_{i,k}$ with $i\neq k$, we quantize both their real and imaginary parts with $B/2\in\{4,2,1\}$ bits of resolution according to \cite{max60}.
In contrast, for the tunable components $Y_{k,k}$, we assume that they can take arbitrary values for simplicity, as they are just $N_T+N_R$.
Comparing Figs.~\ref{fig:tx-sr-noisy-csi} and ~\ref{fig:tx-sr-discrete}, we notice that the impact of discrete-value admittance components with $B\in\{8,4,2\}$ resolution bits is similar to the impact of noisy \gls{csi} with \gls{snr} $\rho\in\{20,10,5\}$~dB, respectively.
This is because the \gls{sqnr} obtained by optimally quantizing a real Gaussian with $B/2\in\{4,2,1\}$ bits is $\rho^\prime\in\{20.2,9.30,4.40\}$~dB \cite{max60}, respectively, which is approximately $\rho\in\{20,10,5\}$~dB.

\section{Conclusion}
\label{sec:conclusion}

In Part~II of this paper, we demonstrate the application of \gls{milac} to wireless communications, highlighting its potential to support communications with thousands of antennas by enabling analog-domain computations and beamforming.
We first derive five special cases of the \gls{lmmse} estimator that have a practical interest in wireless communications, such as the \gls{ls} and \gls{mf}, and illustrate how they can be computed by a \gls{milac} in the analog domain.
Following the capability of a \gls{milac} to compute such special cases, we introduce \gls{milac}-aided beamforming as a novel beamforming strategy enabled by a \gls{milac}.
Compared to state-of-the-art beamforming strategies, \gls{milac}-aided beamforming offers the following five unique benefits in terms of performance, hardware complexity, and computational complexity.
First, it has the same flexibility as digital beamforming, allowing us to achieve maximum performance.
Second, it requires the minimum number of costly \gls{rf} chains, i.e., equal to the number of transmitted symbols.
Third, it can be realized with low-resolution \glspl{adc}/\glspl{dac} since the symbols are precoded fully in the analog domain.
Fourth, it does not perform any operation on a per-symbol basis, since it does not require precoding in the digital domain.
Fifth, it can offer a significant computational complexity reduction in the case of \gls{zf} operations at the transmitter or at the receiver, which can be performed with complexity growing with the square of the antenna number, rather than the cube.
Numerical results support the theoretical insights, showing that \gls{milac}-aided beamforming can offer a remarkable reduction in computational complexity.
For example, \gls{zf} can be achieved with a complexity reduction of $1.5\times10^4$ times.

Future research directions include modeling \gls{milac} accounting for hardware non-idealities and its implementation to assess its feasibility and power consumption.
It will also be interesting to investigate whether \gls{milac} architectures with reduced circuit complexity are viable, for example including a reduced number of tunable admittance components.
Early contributions in this direction can be found in \cite{ner25-4,ner25-5}.

\bibliographystyle{IEEEtran}
\bibliography{IEEEabrv,main}

\begin{thebibliography}{10}
\providecommand{\url}[1]{#1}
\csname url@samestyle\endcsname
\providecommand{\newblock}{\relax}
\providecommand{\bibinfo}[2]{#2}
\providecommand{\BIBentrySTDinterwordspacing}{\spaceskip=0pt\relax}
\providecommand{\BIBentryALTinterwordstretchfactor}{4}
\providecommand{\BIBentryALTinterwordspacing}{\spaceskip=\fontdimen2\font plus
\BIBentryALTinterwordstretchfactor\fontdimen3\font minus \fontdimen4\font\relax}
\providecommand{\BIBforeignlanguage}[2]{{%
\expandafter\ifx\csname l@#1\endcsname\relax
\typeout{** WARNING: IEEEtran.bst: No hyphenation pattern has been}%
\typeout{** loaded for the language `#1'. Using the pattern for}%
\typeout{** the default language instead.}%
\else
\language=\csname l@#1\endcsname
\fi
#2}}
\providecommand{\BIBdecl}{\relax}
\BIBdecl

\bibitem{ner25-2}
M.~Nerini and B.~Clerckx, ``Enabling gigantic {MIMO} beamforming with analog computing,'' \emph{arXiv preprint arXiv:2504.07477v1}, 2025.

\bibitem{ner25-3}
M.~Nerini and B.~Clerckx, ``Analog computing for signal processing and communications -- {Part I}: Computing with microwave networks,'' \emph{arXiv preprint arXiv:2504.06790}, 2025.

\bibitem{lar14}
E.~G. Larsson, O.~Edfors, F.~Tufvesson, and T.~L. Marzetta, ``Massive {MIMO} for next generation wireless systems,'' \emph{IEEE Commun. Mag.}, vol.~52, no.~2, pp. 186--195, 2014.

\bibitem{bjo16}
E.~Björnson, E.~G. Larsson, and T.~L. Marzetta, ``Massive {MIMO}: Ten myths and one critical question,'' \emph{IEEE Commun. Mag.}, vol.~54, no.~2, pp. 114--123, 2016.

\bibitem{bjo24}
E.~Bj{\"o}rnson, F.~Kara, N.~Kolomvakis, A.~Kosasih, P.~Ramezani, and M.~B. Salman, ``Enabling {6G} performance in the upper mid-band by transitioning from massive to gigantic {MIMO},'' \emph{arXiv preprint arXiv:2407.05630}, 2024.

\bibitem{sun14}
S.~Sun, T.~S. Rappaport, R.~W. Heath, A.~Nix, and S.~Rangan, ``{MIMO} for millimeter-wave wireless communications: Beamforming, spatial multiplexing, or both?'' \emph{IEEE Commun. Mag.}, vol.~52, no.~12, pp. 110--121, 2014.

\bibitem{aya14}
O.~E. Ayach, S.~Rajagopal, S.~Abu-Surra, Z.~Pi, and R.~W. Heath, ``Spatially sparse precoding in millimeter wave {MIMO} systems,'' \emph{IEEE Trans. Wireless Commun.}, vol.~13, no.~3, pp. 1499--1513, 2014.

\bibitem{soh16}
F.~Sohrabi and W.~Yu, ``Hybrid digital and analog beamforming design for large-scale antenna arrays,'' \emph{IEEE J. Sel. Top. Signal Process.}, vol.~10, no.~3, pp. 501--513, 2016.

\bibitem{mol17}
A.~F. Molisch, V.~V. Ratnam, S.~Han, Z.~Li, S.~L.~H. Nguyen, L.~Li, and K.~Haneda, ``Hybrid beamforming for massive {MIMO}: A survey,'' \emph{IEEE Commun. Mag.}, vol.~55, no.~9, pp. 134--141, 2017.

\bibitem{ahm18}
I.~Ahmed, H.~Khammari, A.~Shahid, A.~Musa, K.~S. Kim, E.~De~Poorter, and I.~Moerman, ``A survey on hybrid beamforming techniques in {5G}: Architecture and system model perspectives,'' \emph{IEEE Commun. Surv. Tutor.}, vol.~20, no.~4, pp. 3060--3097, 2018.

\bibitem{gon20}
T.~Gong, N.~Shlezinger, S.~S. Ioushua, M.~Namer, Z.~Yang, and Y.~C. Eldar, ``{RF} chain reduction for {MIMO} systems: A hardware prototype,'' \emph{IEEE Syst. J.}, vol.~14, no.~4, pp. 5296--5307, 2020.

\bibitem{dir20}
M.~Di~Renzo, A.~Zappone, M.~Debbah, M.-S. Alouini, C.~Yuen, J.~de~Rosny, and S.~Tretyakov, ``Smart radio environments empowered by reconfigurable intelligent surfaces: How it works, state of research, and the road ahead,'' \emph{IEEE J. Sel. Areas Commun.}, vol.~38, no.~11, pp. 2450--2525, 2020.

\bibitem{wu21-2}
Q.~Wu, S.~Zhang, B.~Zheng, C.~You, and R.~Zhang, ``Intelligent reflecting surface-aided wireless communications: A tutorial,'' \emph{IEEE Trans. Commun.}, vol.~69, no.~5, pp. 3313--3351, 2021.

\bibitem{jam21}
V.~Jamali, A.~M. Tulino, G.~Fischer, R.~R. Müller, and R.~Schober, ``Intelligent surface-aided transmitter architectures for millimeter-wave ultra massive {MIMO} systems,'' \emph{IEEE Open J. Commun. Soc.}, vol.~2, pp. 144--167, 2021.

\bibitem{you22}
C.~You, B.~Zheng, W.~Mei, and R.~Zhang, ``How to deploy intelligent reflecting surfaces in wireless network: {BS}-side, user-side, or both sides?'' \emph{J. Commun. Inf. Netw.}, vol.~7, no.~1, pp. 1--10, 2022.

\bibitem{hua23}
Y.~Huang, L.~Zhu, and R.~Zhang, ``Integrating intelligent reflecting surface into base station: Architecture, channel model, and passive reflection design,'' \emph{IEEE Trans. Commun.}, vol.~71, no.~8, pp. 5005--5020, 2023.

\bibitem{she20}
S.~Shen, B.~Clerckx, and R.~Murch, ``Modeling and architecture design of reconfigurable intelligent surfaces using scattering parameter network analysis,'' \emph{IEEE Trans. Wireless Commun.}, vol.~21, no.~2, pp. 1229--1243, 2022.

\bibitem{li22-1}
H.~Li, S.~Shen, and B.~Clerckx, ``Beyond diagonal reconfigurable intelligent surfaces: From transmitting and reflecting modes to single-, group-, and fully-connected architectures,'' \emph{IEEE Trans. Wireless Commun.}, vol.~22, no.~4, pp. 2311--2324, 2023.

\bibitem{ner23-1}
M.~Nerini, S.~Shen, H.~Li, and B.~Clerckx, ``Beyond diagonal reconfigurable intelligent surfaces utilizing graph theory: Modeling, architecture design, and optimization,'' \emph{IEEE Trans. Wireless Commun.}, vol.~23, no.~8, pp. 9972--9985, 2024.

\bibitem{mis24}
A.~Mishra, Y.~Mao, C.~D’Andrea, S.~Buzzi, and B.~Clerckx, ``Transmitter side beyond-diagonal reconfigurable intelligent surface for massive {MIMO} networks,'' \emph{IEEE Wireless Commun. Lett.}, vol.~13, no.~2, pp. 352--356, 2024.

\bibitem{an23}
J.~An, C.~Xu, D.~W.~K. Ng, G.~C. Alexandropoulos, C.~Huang, C.~Yuen, and L.~Hanzo, ``Stacked intelligent metasurfaces for efficient holographic {MIMO} communications in {6G},'' \emph{IEEE J. Sel. Areas Commun.}, vol.~41, no.~8, pp. 2380--2396, 2023.

\bibitem{an24a}
J.~An, C.~Yuen, C.~Xu, H.~Li, D.~W.~K. Ng, M.~Di~Renzo, M.~Debbah, and L.~Hanzo, ``Stacked intelligent metasurface-aided {MIMO} transceiver design,'' \emph{IEEE Wireless Commun.}, vol.~31, no.~4, pp. 123--131, 2024.

\bibitem{ner24}
M.~Nerini and B.~Clerckx, ``Physically consistent modeling of stacked intelligent metasurfaces implemented with beyond diagonal {RIS},'' \emph{IEEE Commun. Lett.}, vol.~28, no.~7, pp. 1693--1697, 2024.

\bibitem{shl19}
N.~Shlezinger, O.~Dicker, Y.~C. Eldar, I.~Yoo, M.~F. Imani, and D.~R. Smith, ``Dynamic metasurface antennas for uplink massive {MIMO} systems,'' \emph{IEEE Trans. Commun.}, vol.~67, no.~10, pp. 6829--6843, 2019.

\bibitem{wil22}
R.~J. Williams, P.~Ramírez-Espinosa, J.~Yuan, and E.~de~Carvalho, ``Electromagnetic based communication model for dynamic metasurface antennas,'' \emph{IEEE Trans. Wireless Commun.}, vol.~21, no.~10, pp. 8616--8630, 2022.

\bibitem{pro25}
H.~Prod'homme, J.~Tapie, L.~L. Magoarou, and P.~del Hougne, ``Benefits of mutual coupling in dynamic metasurface antennas for optimizing wireless communications--theory and experimental validation,'' \emph{arXiv preprint arXiv:2502.15565}, 2025.

\bibitem{kaw05}
H.~Kawakami and T.~Ohira, ``Electrically steerable passive array radiator ({ESPAR}) antennas,'' \emph{IEEE Antennas Propag. Mag.}, vol.~47, no.~2, pp. 43--50, 2005.

\bibitem{han13}
B.~Han, V.~I. Barousis, C.~B. Papadias, A.~Kalis, and R.~Prasad, ``{MIMO} over {ESPAR} with {16-QAM} modulation,'' \emph{IEEE Wireless Commun. Lett.}, vol.~2, no.~6, pp. 687--690, 2013.

\bibitem{zha23}
C.~Zhang, S.~Shen, Z.~Han, and R.~Murch, ``Analog beamforming using {ESPAR} for single-{RF} precoding systems,'' \emph{IEEE Trans. Wireless Commun.}, vol.~22, no.~7, pp. 4387--4400, 2023.

\bibitem{mul14}
R.~R. M{\"u}ller, M.~A. Sedaghat, and G.~Fischer, ``Load modulated massive {MIMO},'' in \emph{2014 IEEE Global Conference on Signal and Information Processing (GlobalSIP)}.\hskip 1em plus 0.5em minus 0.4em\relax IEEE, 2014, pp. 622--626.

\bibitem{sed16}
M.~A. Sedaghat, V.~I. Barousis, R.~R. Müller, and C.~B. Papadias, ``Load modulated arrays: A low-complexity antenna,'' \emph{IEEE Commun. Mag.}, vol.~54, no.~3, pp. 46--52, 2016.

\bibitem{tan20a}
W.~Tang, M.~Z. Chen, J.~Y. Dai, Y.~Zeng, X.~Zhao, S.~Jin, Q.~Cheng, and T.~J. Cui, ``Wireless communications with programmable metasurface: New paradigms, opportunities, and challenges on transceiver design,'' \emph{IEEE Wireless Commun.}, vol.~27, no.~2, pp. 180--187, 2020.

\bibitem{tan20b}
W.~Tang, J.~Y. Dai, M.~Z. Chen, K.-K. Wong, X.~Li, X.~Zhao, S.~Jin, Q.~Cheng, and T.~J. Cui, ``{MIMO} transmission through reconfigurable intelligent surface: System design, analysis, and implementation,'' \emph{IEEE J. Sel. Areas Commun.}, vol.~38, no.~11, pp. 2683--2699, 2020.

\bibitem{man22}
P.~Mannocci, E.~Melacarne, and D.~Ielmini, ``An analogue in-memory ridge regression circuit with application to massive {MIMO} acceleration,'' \emph{IEEE J. Emerg. Sel. Top. Circuits Syst.}, vol.~12, no.~4, pp. 952--962, 2022.

\bibitem{man23c}
P.~Mannocci, E.~Melacarne, G.~Pedretti, C.~Villa, F.~Sancandi, U.~Spagnolini, and D.~Ielmini, ``Accelerating massive {MIMO} in {6G} communications by analog in-memory computing circuits,'' in \emph{2023 IEEE International Symposium on Circuits and Systems (ISCAS)}, 2023.

\bibitem{wan23}
C.~Wang, G.-J. Ruan, Z.-Z. Yang, X.-J. Yangdong, Y.~Li, L.~Wu, Y.~Ge, Y.~Zhao, C.~Pan, W.~Wei \emph{et~al.}, ``Parallel in-memory wireless computing,'' \emph{Nature Electronics}, vol.~6, no.~5, pp. 381--389, 2023.

\bibitem{oma25b}
Z.~R. Omam, A.~Bagheri, S.~Danesh, S.~E. Hosseininejad, O.~Yurduseven, G.~C. Alexandropoulos, and M.~Khalily, ``{STAR-RIS} for simultaneous communications and fourier-based {AoA} sensing: Design and experimentation,'' \emph{IEEE Antennas Wireless Propag. Lett.}, 2025.

\bibitem{joy25}
A.~T. Joy, A.~Tishchenko, H.~Taghvaee, P.~Mursia, V.~Sciancalepore, and M.~Khalily, ``{RIS}-enabled {ISAC} in {6G}: Exploring the role of wave domain computing,'' \emph{IEEE Commun. Stand. Mag.}, 2025.

\bibitem{an24b}
J.~An, C.~Yuen, Y.~L. Guan, M.~Di~Renzo, M.~Debbah, H.~Vincent~Poor, and L.~Hanzo, ``Two-dimensional direction-of-arrival estimation using stacked intelligent metasurfaces,'' \emph{IEEE J. Sel. Areas Commun.}, vol.~42, no.~10, pp. 2786--2802, 2024.

\bibitem{gol13}
M.~Goldenbaum, H.~Boche, and S.~Stańczak, ``Harnessing interference for analog function computation in wireless sensor networks,'' \emph{IEEE Trans. Signal Process.}, vol.~61, no.~20, pp. 4893--4906, 2013.

\bibitem{zhu19}
G.~Zhu and K.~Huang, ``{MIMO} over-the-air computation for high-mobility multimodal sensing,'' \emph{IEEE Internet Things J.}, vol.~6, no.~4, pp. 6089--6103, 2019.

\bibitem{zhu21}
G.~Zhu, J.~Xu, K.~Huang, and S.~Cui, ``Over-the-air computing for wireless data aggregation in massive {IoT},'' \emph{IEEE Wireless Commun.}, vol.~28, no.~4, pp. 57--65, 2021.

\bibitem{yan20}
K.~Yang, T.~Jiang, Y.~Shi, and Z.~Ding, ``Federated learning via over-the-air computation,'' \emph{IEEE Trans. Wireless Commun.}, vol.~19, no.~3, pp. 2022--2035, 2020.

\bibitem{cle13}
B.~Clerckx and C.~Oestges, \emph{{MIMO} wireless networks: Channels, techniques and standards for multi-antenna, multi-user and multi-cell systems}.\hskip 1em plus 0.5em minus 0.4em\relax Academic Press, 2013.

\bibitem{bal15}
C.~A. Balanis, \emph{Antenna theory: Analysis and design}.\hskip 1em plus 0.5em minus 0.4em\relax John Wiley \& Sons, 2015.

\bibitem{joh07}
S.~G. Johnson and M.~Frigo, ``A modified split-radix {FFT} with fewer arithmetic operations,'' \emph{IEEE Trans. Signal Process.}, vol.~55, no.~1, pp. 111--119, 2007.

\bibitem{max60}
J.~Max, ``Quantizing for minimum distortion,'' \emph{IRE Trans. Inf. Theory}, vol.~6, no.~1, pp. 7--12, 1960.

\bibitem{ner25-4}
M.~Nerini and B.~Clerckx, ``Capacity of {MIMO} systems aided by microwave linear analog computers ({MiLACs}),'' \emph{arXiv preprint arXiv:2506.05983}, 2025.

\bibitem{ner25-5}
M.~Nerini and B.~Clerckx, ``{MIMO} systems aided by microwave linear analog computers: Capacity-achieving architectures with reduced circuit complexity,'' \emph{arXiv preprint arXiv:2506.15052}, 2025.

\end{thebibliography}

\end{document}